\documentclass[reqno,12pt]{amsart}
\usepackage{fullpage}
\usepackage{amsfonts}
\usepackage{amssymb}

\numberwithin{equation}{section}
\def\X{{\bf X}}
\def\pr{{\rm pr}}
\def\p{\partial}

\def\E{{\mathcal E}}
\def\triv{{\rm triv}}

\def\const{{\rm const.}}
\def\Rnum{\mathbb{R}}

\def\const{{\rm const.}}

\def\Dop{{\mathcal D}}
\def\t{{\rm t}}

\def\parder#1#2{\frac{\partial{#1}}{\partial{#2}}}
\def\parderop#1{\partial/\partial{#1}}

\def\up#1{{}^{\scriptstyle #1}}

\newtheorem{prop}{Proposition}
\newtheorem{thm}{Theorem}
\newtheorem{cor}{Corollary}
\newtheorem{lem}{Lemma}

\def\propref#1{Proposition~\ref{#1}}
\def\thmref#1{Theorem~\ref{#1}}
\def\lemref#1{Lemma~\ref{#1}}
\def\Ref#1{Ref.\cite{#1}}
\def\Refs#1{Refs.\cite{#1}}
\def\secref#1{Sec.~\ref{#1}}

\def\ie/{i.e.}
\def\eg/{e.g.}
\def\etc/{etc.}

\begin{document}
\allowdisplaybreaks[3]

\title{Symmetry-invariant conservation laws of\\
partial differential equations}

\author{
Stephen C. Anco$^1$
\lowercase{\scshape{and}}
Abdul H. Kara$^2$\\
\\\lowercase{\scshape{
${}^1$Department of Mathematics and Statistics\\
Brock University\\
St. Catharines, ON L2S3A1, Canada}} \\\\
\lowercase{\scshape{
${}^2$School of Mathematics\\
University of the Witwatersrand\\
Wits 2050, Johannesburg, South Africa}}\\
}

\begin{abstract}
A simple characterization of the action of symmetries on conservation laws of
partial differential equations is studied by using the general method of 
conservation law multipliers. 
This action is used to define symmetry-invariant and symmetry-homogeneous conservation laws. 
The main results are applied to several examples of physically interest, 
including the generalized Korteveg-de Vries equation, 
a non-Newtonian generalization of Burger's equation, 
the $b$-family of peakon equations, 
and the Navier-Stokes equations for compressible, viscous fluids in two dimensions. 
\end{abstract}

\maketitle

\section{Introduction}
\label{intro}

Conservation laws have several important uses 
in the study of partial differential equations (PDEs),
especially for determining conserved quantities and constants of motion, 
detecting integrability and linearizations, 
finding potentials and nonlocally-related systems,
as well as checking the accuracy of numerical solution methods. 

For any PDE system, regardless of whether a Lagrangian exists, 
the conservation laws admitted by the system 
can be found by a direct computational method \cite{AncBlu97,AncBlu02a,AncBlu02b}
which is similar to Lie's method 
for finding the symmetries \cite{Olv,1stbook,2ndbook} admitted by the system. 
The method for conservation laws uses a characteristic form involving multipliers 
and is based on the standard result \cite{Alo,Vin,Olv,Anc-review} 
that there is a direct correspondence between conservation laws and multipliers
whenever a PDE system can be expressed in a solved form 
in terms of a set of leading derivatives. 
This is the same property used by Lie's method for symmetries. 
In particular, 
all multipliers can be found by solving 
a linear system of determining equations which consists of 
the adjoint of the symmetry determining equations \cite{Vin,Olv,AncBlu97,AncBlu02a,AncBlu02b,Anc-review}
plus additional equations analogous to Helmholtz conditions \cite{Vin,AncBlu97,AncBlu02a,AncBlu02b,Anc-review}. 
For each solution of this determining system, 
a corresponding conservation law can be obtained 
by various direct integration methods \cite{2ndbook,Anc-review}. 
Consequently, 
the problem of finding all conservation laws is reduced to a kind of 
adjoint of the problem of finding all symmetries. 

In the case when a PDE system has a Lagrangian formulation, 
the symmetry determining equations constitute a self-adjoint linear system,
and the multiplier determining system becomes the same as 
the determining equations for variational symmetries 
\cite{Vin,Olv,2ndbook,BluCheAnc}. 
Hence, 
the conservation laws admitted by a Lagrangian PDE system can be found 
without the need to use the Lagrangian 
(and without the awkward need to consider ``gauge terms'' 
in the definition of variational symmetries). 

For any given PDE system, 
the admitted symmetries have a natural action 
on the admitted conservation laws 
\cite{Olv,Ibr,Kha,IbrKarMah,BluTemAnc}. 
This action allows conservation laws to be divided into 
symmetry equivalence classes, 
which can be used to define a generating subset (or a basis) \cite{Ibr,Kha,KarMah02} 
for the set of all conservation laws of the PDE system. 
The study of conservation laws that are invariant under the action of 
a given set of admitted symmetries was introduced in \Ref{KarMah00}. 
More recent work \cite{Anc16a} has explored the properties of 
symmetry-invariant conservation laws, 
as well as conservation laws that are, more generally, homogeneous 
(mapped into themselves) 
under the action of symmetries. 
Such conservation laws have at least three interesting applications \cite{Anc16a}. 
Firstly, 
each symmetry-homogeneous conservation law 
represents a one-dimensional invariant subspace 
in the set of all non-trivial conservation laws. 
This is a useful feature when a generating subset (or basis) is being sought.
Secondly, 
any symmetry-invariant conservation law will reduce to a first integral 
for the ODE obtained by symmetry reduction of the given PDE system 
when symmetry-invariant solutions of the system are sought. 
This provides a direct reduction of order of the ODE. 
Thirdly, 
the determining equations for multipliers can be augmented by 
a symmetry-homogeneity condition,
which allows symmetry-homogeneous conservation laws to be obtained 
in a direct way by solving the augmented determining system. 

In the present paper, 
we further develop this work
and apply it to several nonlinear PDE systems 
arising in a variety of physical applications,
including water waves, fluid flow, and gas dynamics. 

In \secref{prelim}, 
the multiplier method for finding the conservation laws of a given PDE system
is reviewed, along with the action of symmetries on conservation laws. 
In \secref{results}, 
the main properties of symmetry-invariant conservation laws and
symmetry-homogeneous conservation laws are derived 
for PDE systems in any number of variables,
generalizing the results introduced in \Ref{Anc16a}
for single PDEs with two independent variables. 
Examples and applications of the multiplier method 
and of the general results on symmetry properties for conservation laws 
are presented in \secref{examples}. 
Finally, some concluding remarks are made in \secref{remarks}.

\section{Preliminaries}
\label{prelim}

Consider an $N$th-order system of PDEs 
\begin{equation}\label{pde}
G^a(t,x,u,\p u,\ldots,\p^N u)=0,\ 
a=1,\ldots, M
\end{equation}
with $m$ dependent variables $u^\alpha$, $\alpha=1,\ldots,m$, 
and $n+1$ independent variables $t,x^i$, $i=1,\ldots,n$, 
where $\p^k u$ denotes all $k$th order partial derivatives of $u$ with respect to $t$ and $x^i$. 
A partial derivative of $u$ in the PDE system \eqref{pde} is a {\em leading derivative} 
\cite{Sch98,Sch07}
if no differential consequences of it appear in the system. 
The PDE system \eqref{pde} will be called {\em normal} 
if the $M$ PDEs can be expressed in a solved form in terms of 
a set of leading derivatives of $u$ with respect to a single independent variable
(after a point transformation if necessary)
such that the right-hand side of each PDE does not contain 
any of these derivatives and their differential consequences. 
As seen in the examples in \secref{examples}, 
typical PDE systems (without differential identities) 
arising in physical applications are normal. 

A {\em conservation law} \cite{Olv,2ndbook} of a given PDE system \eqref{pde} 
is a total space-time divergence expression that vanishes 
on the solution space $\E$ of the system,
\begin{equation}\label{conslaw}
(D_t T(t,x,u,\p u,\ldots,\p^r u) +D_i X^i(t,x,u,\p u,\ldots,\p^r u))|_\E=0 . 
\end{equation}
Physically, this is a local continuity equation, 
where the function $T$ is a conserved density 
and the vector function $X^i$ is a spatial flux. 
The pair 
\begin{equation}\label{current}
(T,X^i)=\Phi 
\end{equation}
is called a {\em conserved current}. 
Throughout, 
\begin{equation}
D_t =\parderop{t} + u^\alpha_t\parderop{u^\alpha} + u^\alpha_{tx\up{i}}\parderop{u^\alpha_{x\up{i}}} + u^\alpha_{tt}\parderop{u^\alpha_t} +\cdots 
\end{equation}
denotes a total $t$-derivative, 
and 
\begin{equation}
D_i =\parderop{x^i} + u^\alpha_{x\up{i}}\parderop{u^\alpha} + u^\alpha_{tx\up{i}}\parderop{u^\alpha_t} + u^\alpha_{x\up{i} x\up{j}}\parderop{u^\alpha_{x\up{j}}} +\cdots 
\end{equation}
denotes a total $x^i$-derivative. 
Also, the notation $f|_\E$ for a function $f(t,x,u,\p u,\ldots,\p^r u)$ 
means that the given PDE system \eqref{pde} 
as well as its differential consequences are to be used in the evaluation of $f$ 
on the solution space $\E$. 

Every conservation law \eqref{conslaw} can be integrated over 
any given spatial domain $\Omega\subseteq\Rnum^n$
\begin{equation}
\frac d{dt} \int_{\Omega} T d^nx = -\int_{\p\Omega} X^i dA_i
\end{equation}
showing that the rate of change of the quantity 
\begin{equation}\label{C}
\mathcal C[u]= \int_{\Omega} T d^nx
\end{equation}
is balanced by the net flux through the domain boundary $\p\Omega$. 
Here $dA_i$ denotes the area element multiplied by the unit outward normal vector on $\p\Omega$. 

Two conservation laws are locally equivalent if they give the same 
conserved quantity \eqref{C} up to boundary terms. 
This occurs iff their conserved densities
differ by a total spatial divergence $D_i\Theta^i$ on the solution space $\E$,
and correspondingly, 
their fluxes differ by a total time derivative $-D_t\Theta^i$ 
modulo a divergence-free vector $D_j\Gamma^{ij}$,
where $\Theta^i(t,x,u,\p u,\ldots,\p^r u)$ is some vector function
and $\Gamma^{ij}(t,x,u,\p u,\ldots,\p^r u)$ is some antisymmetric tensor function
on $\E$. 
A conservation law is thereby called {\em locally trivial} if 
\begin{equation}\label{trivconslaw}
T|_\E = D_i\Theta^i|_\E,
\quad
X^i|_\E = - D_t\Theta^i|_\E +D_j\Gamma^{ij}|_\E, 
\end{equation}
so then any two locally equivalent conservation laws differ by 
a trivial conservation law. 
For a given PDE system \eqref{pde}, 
the set of all non-trivial conservation laws (up to equivalence)
forms a vector space on which the symmetries of the system have a natural action. 

An {\em infinitesimal symmetry} \cite{Olv,1stbook,2ndbook} 
of a given PDE system \eqref{pde} is a generator 
\begin{equation}\label{symm}
\X=\tau(t,x,u,\p u,\ldots,\p^r u)\parderop{t} +\xi^i(t,x,u,\p u,\ldots,\p^r u)\parderop{x^i} +\eta^\alpha(t,x,u,\p u,\ldots,\p^r u)\parderop{u^\alpha}
\end{equation}
whose prolongation $\pr\X$ leaves invariant the PDE system, 
\begin{equation}
\pr\X(G^a)|_\E =0 . 
\end{equation}
When acting on solutions $u^\alpha(t,x)$ of the PDE system, 
any infinitesimal symmetry \eqref{symm} is equivalent to 
a generator with the {\em characteristic form} 
\begin{equation}\label{symmchar}
\hat\X=P^\alpha\parderop{u^\alpha}, 
\quad 
P^\alpha =\eta^\alpha-\tau u^\alpha_t-\xi^i u^\alpha_{x\up{i}}
\end{equation}
where the characteristic functions $\eta^\alpha$, $\tau$, $\xi^i$ 
are determined by 
\begin{equation}\label{symmdeteq}
0 = \pr\hat\X(G^a)|_\E = (\delta_P G)^a|_\E . 
\end{equation}
Throughout, 
\begin{equation}\label{frechet}
\delta_g f = 
\parder{f}{u^\alpha} g^\alpha 
+ \parder{f}{u^\alpha_t}D_t g^\alpha + \parder{f}{u^\alpha_{x\up{i}}}D_i g^\alpha 
+ \parder{f}{u^\alpha_{tt}}D_t{}^2 g^\alpha 
+ \parder{f}{u^\alpha_{tx\up{i}}}D_tD_i g^\alpha 
+ \parder{f}{u^\alpha_{x\up{i}x\up{j}}}D_iD_j g^\alpha 
+ \cdots
\end{equation}
denotes the Frechet derivative with respect to $u$, 
for any differential functions $f(t,x,u,\p u,\p^2 u,\ldots)$
and $g^\alpha(t,x,u,\p u,\p^2 u,\ldots)$. 

An infinitesimal symmetry of a given PDE system \eqref{pde} is {\em trivial} 
if its action on the solution space $\E$ of the system is trivial, 
$\hat\X u^\alpha =0$ for all solutions $u^\alpha(t,x)$. 
This occurs iff $P^\alpha|_\E=0$. 
The corresponding generator \eqref{symm} of a trivial symmetry 
is thus given by 
$\X|_\E= \tau|_\E\parderop{t} + \xi^i|_\E\parderop{x^i} +(\tau u^\alpha_t +\xi^i u^\alpha_{x\up{i}})|_\E\parderop{u^\alpha}$, 
which has the prolongation $\pr\X|_\E=\tau|_\E D_t + \xi^i|_\E D_i$. 
Conversely, any generator of this form on the solution space $\E$ 
determines a trivial symmetry. 

The {\em differential order of an infinitesimal symmetry} is defined to be 
the maximal differential order among its characteristic functions $P^\alpha|_\E$
evaluated on the solution space $\E$. 
Point symmetries are singled out by having the characteristic form \cite{Olv,1stbook,2ndbook} 
\begin{equation}\label{pointsymm}
P^\alpha=\eta^\alpha(t,x,u) - \tau(t,x,u) u^\alpha_t -\xi^i(t,x,u) u^\alpha_{x\up{i}} 
\end{equation}
which generates a transformation group on $(t,x^i,u^\alpha)$. 
Symmetries having a more general first-order characteristic form 
\begin{equation}\label{1stordsymm}
P^\alpha=\hat\eta^\alpha(t,x,u,\p u)
\end{equation}
do not generate a transformation group \cite{Olv,1stbook} 
unless the number of dependent variables is $m=1$, 
in which case the transformation group is a contact symmetry 
acting on $(t,x^i,u,u_t,u_{x\up{i}})$. 

The action of an infinitesimal symmetry \eqref{symm} 
on the set of conserved currents \eqref{current} is given by 
\cite{Olv,Ibr,Kha,IbrKarMah,BluTemAnc,Anc16a} 
\begin{equation}
T_\X = \pr\X(T) + TD_i\xi^i - X^iD_i\tau,
\quad
X^i_\X = \pr\X(X^i) + X^i(D_t\tau+D_iX^i) -X^jD_j\xi^i-TD_t\xi^i. 
\end{equation}
When the symmetry is expressed in the characteristic form \eqref{symmchar}, 
its action has the simple form 
\begin{equation}\label{symmactionTX}
T_{\hat\X} = \pr\hat\X(T)=\delta_P T, 
\quad
X^i_{\hat\X} = \pr\hat\X(X^i)=(\delta_P X)^i . 
\end{equation}
The conserved currents $(T_\X,X^i_\X)$ and $(T_{\hat\X},X^i_{\hat\X})$ are locally equivalent, 
\begin{equation}
(T_{\hat\X}-T_\X)|_\E = D_i\Theta^i|_\E,
\quad
(X^i_{\hat\X} - X^i_\X)|_\E = -D_t\Theta^i|_\E + D_j\Gamma^{ij}|_\E,
\end{equation}
with 
\begin{equation}
\Theta^i =\tau X^i -T\xi^i,
\quad
\Gamma^{ij} = \xi^iX^j-\xi^jX^i
\end{equation}
which follows from the relation $\pr\X-\pr\hat\X=\tau D_t + \xi^i D_i$. 

An important question is 
when does the symmetry action on a given conserved current 
produce a trivial conserved current?
A simple necessary and sufficient condition can be formulated 
by using a characteristic (canonical) form 
\cite{Alo,Olv,AncBlu97,AncBlu02a,AncBlu02b,2ndbook} for conservation laws,
based on the following standard results \cite{Vin,Olv,Anc-review}. 

\begin{lem}\label{hadamard}
If a differential function $f(t,x,u,\p u,\p^2 u,\ldots)$ vanishes 
on the solution space $\E$ of a given PDE system \eqref{pde}, 
then $f= R_f(G)$ holds identically,
where 
\begin{equation}
R_f =
R_f^{(0)}{}_a + R_f^{(1)}{}_a D_t + R_f^{(1)}{}^i_a D_i 
+ R_f^{(2)}{}_a D_t{}^2 + R_f^{(2)}{}^i_a D_tD_i + R_f^{(2)}{}^{ij}_a D_iD_j +\cdots 
\end{equation}
is a linear differential operator, depending on $f$, 
with coefficients that are non-singular on $\E$ whenever the PDE system is normal. 
\end{lem}

\begin{lem}\label{totaldiv} 
A differential function $f(t,x,u,\p u,\p^2 u,\ldots)$ is a total space-time divergence $f=D_t A +D_i B^i$ for some functions
$A(t,x,u,\p u,\p^2 u,\ldots)$ and $B^i(t,x,u,\p u,\p^2 u,\ldots)$ 
iff $E_{u\up{\alpha}}(f)=0$ holds identically,
where 
\begin{equation}
E_{u\up{\alpha}} = 
\parderop{u^\alpha} -D_t\parderop{u^\alpha_t}-D_i\parderop{u^\alpha_{x\up{i}}} 
+ D_t{}^2\parderop{u^\alpha_{tt}} + D_tD_i\parderop{u^\alpha_{tx\up{i}}} + D_iD_j\parderop{u^\alpha_{x\up{i}x\up{j}}} 
+\cdots
\end{equation}
is the Euler-Lagrange operator. 
\end{lem}

From \lemref{hadamard},
a conservation law $(D_t T +D_i X^i)|_\E=0$ for a normal PDE system $G^a=0$
can be expressed as a divergence identity
\begin{equation}\label{conslawoffE}
D_t T +D_i X^i = R_\Phi(G)
\end{equation}
holding off of the solution space $\E$ of the PDE system, 
where $u^\alpha(t,x)$ is an arbitrary (sufficiently smooth) function. 
In this identity, 
integration by parts on the terms in $R_\Phi(G)$ yields 
\begin{equation}\label{chareqn}
D_t\tilde T +D_i\tilde X^i= Q_aG^a
\end{equation}
with 
\begin{equation}\label{equivTX}
\begin{aligned}
\tilde T  & = 
T + R_\Phi^{(1)}{}_aG^a + R_\Phi^{(2)}{}_aD_t G^a -(D_tR_\Phi^{(2)}{}_a)G^a + R_\Phi^{(2)}{}^i_aD_i G^a + \cdots,
\\\tilde X^i & = 
X^i + R_\Phi^{(1)}{}^i_aG^a + R_\Phi^{(2)}{}^{ij}_aD_j G^a -(D_jR_\Phi^{(2)}{}^{ij})G^a - R_\Phi^{(2)}{}^{i}D_t G + \cdots, 
\end{aligned}
\end{equation}
and
\begin{equation}\label{Q}
Q_a = R_\Phi^{(0)}{}_a - D_t R_\Phi^{(1)}{}_a -D_i R_\Phi^{(1)}{}^i_a +D_t{}^2R_\Phi^{(2)}{}_a +D_tD_iR_\Phi^{(2)}{}^i_a +D_iD_jR_\Phi^{(2)}{}^{ij}_a + \cdots . 
\end{equation}
On the solution space $\E$ of the PDE system $G^a=0$, 
note that $\tilde T|_\E = T|_\E$ and $\tilde X^i|_\E = X^i|_\E$ reduce to 
the conserved density and the flux in the conservation law 
$(D_t T +D_i X^i)|_\E=0$, 
and hence $(D_t\tilde T +D_i\tilde X^i)|_\E= 0$ is a locally equivalent conservation law.
The identity \eqref{chareqn} is called the {\em characteristic equation}
for the conservation law, 
and the set of functions \eqref{Q} is called the {\em multiplier}. 
In general, 
a set of functions $Q^a(t,x,u,\p u,\p^2 u,\ldots\p^r u)$, $a=1,\ldots,M$, 
will be a multiplier iff its summed product with the PDEs $G^a=0$, $a=1,\ldots,M$, 
has the form of a total space-time divergence. 

From the characteristic equation \eqref{chareqn}, 
it is straightforward to prove \cite{Vin,Olv,Anc-review} the following basic result, 
which underlies the generality of the multiplier method. 

\begin{thm}\label{correspondence}
For normal PDE systems \eqref{pde} (with no differential identities), 
there is a one-to-one correspondence between conservation laws (up to equivalence) and multipliers evaluated on the solution space of the system. 
\end{thm}

A generalized version of this correspondence can be shown to hold 
for PDE systems that possess differential identities, 
such as Maxwell's electromagnetic field equations, 
magnetohydrodynamic systems, 
and Einstein's gravitational field equations.
(See \Ref{Ver,Anc-review}.)

The {\em differential order of a conservation law} is defined to be 
the smallest differential order among all locally equivalent conserved currents.
For any normal PDE system, 
a conservation law is said to be of {\em low order} 
if each derivative variable $\p^k u^\alpha$ that appears in its multiplier 
is related to a leading derivative of $u^\alpha$ 
by differentiations with respect to $t,x^i$. 
(Note that, therefore, the differential order $r$ of the multiplier 
must be strictly less than the differential order $N$ of the PDE system.)
As seen in the examples in \secref{examples}, 
physically important conservation laws, such as energy and momentum, 
are always of low order,
whereas higher order conservation laws are typically connected with integrability. 

All conservation law multipliers for any normal PDE system 
can be determined from \lemref{totaldiv} applied to 
the characteristic equation \eqref{chareqn}. 
This yields 
\begin{equation}\label{multdeteq}
0=E_{u^\alpha}(Q_aG^a)=(\delta_Q^* G)_\alpha +(\delta_G^* Q)_\alpha
\end{equation}
which is required to hold identically.
Here, a star denotes the adjoint of the Frechet derivative with respect to $u$:
\begin{equation}\label{adjfrechet}
\begin{aligned}
(\delta_g^* f)_\alpha & = 
\parder{f}{u^\alpha} g
-D_t\Big(\parder{f}{u^\alpha_t} g\Big) -D_i\Big(\parder{f}{u^\alpha_x\up{i}}g\Big)
+ D_t{}^2\Big(\parder{f}{u^\alpha_{tt}}g\Big)
\\&\qquad
+ D_tD_i\Big(\parder{f}{u^\alpha_{tx\up{i}}}g\Big)
+ D_iD_j\Big(\parder{f}{u^\alpha_{x\up{i}x\up{j}}}g\Big)
+ \cdots
\\
& = 
E_{u\up{\alpha}}(f)g -E_{u\up{\alpha}}^{(1,t)}(f)D_t g -E_{u\up{\alpha}}^{(1,i)}(f)D_i g
+ E_{u\up{\alpha}}^{(2,tt)}(f)D_t{}^2 g 
\\&\qquad
+E_{u\up{\alpha}}^{(2,ti)}(f)D_tD_i g
+E_{u\up{\alpha}}^{(2,ij)}(f)D_iD_j g 
+ \cdots
\end{aligned}
\end{equation}
for any differential functions $f(t,x,u,\p u,\p^2 u,\ldots)$
and $g(t,x,u,\p u,\p^2 u,\ldots)$,
where $E_{u\up{\alpha}}^{(1,t)}$, $E_{u\up{\alpha}}^{(1,i)}$, 
$E_{u\up{\alpha}}^{(2,tt)}$, $E_{u\up{\alpha}}^{(2,ti)}$, $E_{u\up{\alpha}}^{(2,ij)}$, 
\etc/ denote higher Euler operators \cite{Olv}. 
The determining equation \eqref{multdeteq} 
can be converted into a linear system of equations for $Q_a$ 
by the following steps. 

On the solution space $\E$ of a given PDE system \eqref{pde},
the Frechet derivative operators $\delta_G|_\E$ and $\delta_G^*|_\E$ vanish. 
Thus, the determining equation \eqref{multdeteq} implies
\begin{equation}\label{adjsymmdeteq}
(\delta_Q^* G)_\alpha|_\E =0 . 
\end{equation}
From \lemref{hadamard}, it then follows that $Q_a$ satisfies the identity 
\begin{equation}\label{adjsymmid}
(\delta_Q^* G)_\alpha = R_Q(G)_\alpha 
\end{equation}
for some linear differential operator 
\begin{equation}\label{RQop}
(R_Q)_\alpha =
R_Q^{(0)}{}_{a\alpha} + R_Q^{(1)}{}_{a\alpha} D_t + R_Q^{(1)}{}^i_{a\alpha} D_i 
+ R_Q^{(2)}{}_{a\alpha} D_t{}^2 + R_Q^{(2)}{}^i_{a\alpha} D_tD_i + R_Q^{(2)}{}^{ij}_{a\alpha} D_iD_j +\cdots 
\end{equation}
whose coefficients are non-singular on $\E$ 
as the PDE system $G^a=0$ is assumed to be normal. 
Substitution of this identity \eqref{adjsymmid} 
into the determining equation \eqref{multdeteq} 
yields 
\begin{equation}\label{helmholtzoffE}
0=R_Q(G)_\alpha+ (\delta_G^* Q)_\alpha
\end{equation}
where $u^\alpha(t,x)$ is an arbitrary (sufficiently smooth) function. 
This equation \eqref{helmholtzoffE} can be split with respect to 
the set of leading derivatives of $u^\alpha$ in the normal form of the PDE system
and each differential consequence of these derivatives. 
The splitting yields a linear system of equations:
\begin{subequations}\label{helmholtzeq}
\begin{align}
& R_Q^{(0)}{}_{a\alpha} = E_{u\up{\alpha}}(Q_a) ,
\\
& R_Q^{(1)}{}_{a\alpha} = -E_{u\up{\alpha}}^{(1,t)}(Q_a),
\quad
R_Q^{(1)}{}^i_{a\alpha} = -E_{u\up{\alpha}}^{(1,i)}(Q_a),
\\
& R_Q^{(2)}{}_{a\alpha} = E_{u\up{\alpha}}^{(2,tt)}(Q_a),
\quad
R_Q^{(2)}{}^i_{a\alpha}  = E_{u\up{\alpha}}^{(2,ti)}(Q_a),
\quad
R_Q^{(2)}{}^{ij}_{a\alpha}  = E_{u\up{\alpha}}^{(2,ij)}(Q_a),
\\
&\qquad\vdots
\nonumber
\end{align}
\end{subequations}
Hence, we obtain the following determining system for conservation law multipliers. 

\begin{prop}
The determining equation \eqref{multdeteq} for conservation law multipliers 
of a normal PDE system \eqref{pde} 
is equivalent to the linear system of equations \eqref{adjsymmdeteq} and \eqref{helmholtzeq}. 
The first equation \eqref{adjsymmdeteq} is 
the adjoint of the symmetry determining equation \eqref{symmdeteq},
and its solutions $Q_a(t,x,u,\p u,\p^2 u,\ldots)$ 
are called {\em adjoint-symmetries} \cite{AncBlu97,AncBlu02a,AncBlu02b}. 
The remaining equations \eqref{helmholtzeq} comprise Helmholtz conditions \cite{Olv,Anc-review}
which are necessary and sufficient for an adjoint-symmetry $Q_a(t,x,u,\p u,\p^2 u,\ldots)$ 
to have the variational form \eqref{Q} where $\Phi=(T,X^i)$ is a conserved current. 
\end{prop}

These equations can be solved computationally 
by the same standard procedure \cite{Olv,1stbook,2ndbook}
used to solve the determining equation for symmetries. 
Moreover, the multiplier determining system is, in general, 
more overdetermined than is the symmetry determining equation,
and hence the computation of multipliers is generally easier 
than the computation of symmetries. 

This formulation of a determining system for conservation law multipliers 
has a simple adjoint relationship to Noether's theorem, 
as explained in \Ref{Anc-review}. 
First, recall \cite{Olv,1stbook,2ndbook} that the condition 
for a PDE system \eqref{pde} to be given by Euler-Lagrange equations 
\begin{equation}\label{ELpde}
G^a=E_{u\up{\alpha}}(L)=0
\end{equation}
for some Lagrangian $L(t,x,u,\p u,\p^2 u,\ldots)$ 
is that 
\begin{equation}\label{ELpde-deteqn}
(\delta_f G)^a=(\delta_f^* G)_\alpha
\end{equation}
holds for arbitrary differential functions $f^\alpha(t,x,u,\p u,\p^2 u,\ldots)$.
In particular, 
it is necessary that in the PDE system 
the number $M$ of equations is the same as the number $m$ of dependent variables, 
whereby the indices $a$ and $\alpha$ can be identified. 

The following result is now straightforward to establish 
(see \Refs{Vin,AncBlu97,AncBlu02a,AncBlu02b,Anc-review}).

\begin{thm}\label{adjnoether}
For any normal PDE system \eqref{pde}, 
conservation law multipliers are adjoint-symmetries \eqref{adjsymmdeteq}
that satisfy Helmholtz conditions \eqref{helmholtzeq}. 
In the case when the PDE system is an Euler-Lagrange system \eqref{ELpde}, 
adjoint-symmetries are the same as symmetries,
and the Helmholtz conditions are equivalent to symmetry invariance of the
Lagrangian modulo a total space-time divergence,
so multipliers for a Lagrangian PDE system 
consequently are the same as variational symmetries.
\end{thm}

In applications of \thmref{adjnoether}, 
the use of a Lagrangian to obtain the conserved current from a variational symmetry 
is replaced by 
either a homotopy integral formula \cite{Olv,AncBlu02a,AncBlu02b} 
or direct integration of the characteristic equation \cite{Wol02b,2ndbook},
both of which are applicable for any normal PDE system. 
If a given PDE system possesses a scaling symmetry, 
then any conserved current having non-zero scaling weight 
can be obtained from an algebraic formula \cite{Anc03} in terms of a multiplier
(see also \Refs{DecNiv,PooHer}).

\section{Main results}
\label{results}

A simple expression for the action of an infinitesimal symmetry 
on a conservation law multiplier will now be presented. 
The following Frechet derivative identity \cite{Cav,Lun,Zha86,AncBlu96,AncBlu97,AncBlu02a,AncBlu02b}
is needed. 

\begin{lem}
The Frechet derivative \eqref{frechet} and its adjoint \eqref{adjfrechet}
satisfy the identity 
\begin{equation}\label{frechetid}
h\delta_g f- g\delta_h^* f = D_t\Psi_f^t(g,h) + D_i\Psi_f^i(g,h)
\end{equation}
with
\begin{equation}
\begin{aligned}
\Psi_f^t(g,h) = & 
g^\alpha\Big( h \parder{f}{u^\alpha_t} -D_t\Big(h \parder{f}{u^\alpha_{tt}}\Big) -D_i\Big(h \parder{f}{u^\alpha_{tx\up{i}}}\Big) +\cdots \Big)
\\&\qquad
+ D_t g^\alpha\Big( h \parder{f}{u^\alpha_{tt}} -D_t\Big(h \parder{f}{u^\alpha_{ttt}}\Big) -D_i\Big(h \parder{f}{u^\alpha_{ttx\up{i}}}\Big) +\cdots \Big)
\\&\qquad
+ D_t{}^2 g^\alpha\Big( h \parder{f}{u^\alpha_{ttt}} -D_t\Big(h \parder{f}{u^\alpha_{tttt}}\Big) -D_i\Big(h \parder{f}{u^\alpha_{tttx\up{i}}}\Big) +\cdots \Big)
+\cdots
\end{aligned}
\end{equation}
modulo a trivial term $D_i\Theta^i$,
and
\begin{equation}
\begin{aligned}
\Psi_f^i(g,h) = & 
g^\alpha\Big( h\parder{f}{u^\alpha_{x\up{i}}} -D_j\Big(h\parder{f}{u^\alpha_{x\up{i}x\up{j}}}\Big) +D_jD_k\Big(h \parder{f}{u^\alpha_{x\up{i}x\up{j}x\up{k}}}\Big) +\cdots )
\\&\qquad
+ D_t g^\alpha\Big( h \parder{f}{u^\alpha_{tx\up{i}}} -D_j\Big(h \parder{f}{u^\alpha_{tx\up{i}x\up{j}}}\Big) +D_jD_k\Big(h\parder{f}{u^\alpha_{tx\up{i}x\up{j}x\up{k}}}\Big) +\cdots \Big)
\\&\qquad
+ D_j g^\alpha\Big( h \parder{f}{u^\alpha_{x\up{i}x\up{j}}} -D_k\Big(h \parder{f}{u^\alpha_{x\up{i}x\up{j}x\up{k}}}\Big) +D_kD_l\Big(h \parder{f}{u^\alpha_{x\up{i}x\up{j}x\up{k}x\up{l}}}\Big) +\cdots \Big)
+\cdots
\end{aligned}
\end{equation}
modulo a trivial term $-D_t\Theta^i+D_j\Gamma^{ij}$, 
where $f(t,x,u,\p u,\p^2 u,\ldots)$, $g^\alpha(t,x,u,\p u,\p^2 u,\ldots)$ and $h(t,x,u,\p u,\p^2 u,\ldots)$ 
are arbitrary differential functions. 
\end{lem}

As shown independently in \Ref{Cav,Lun,Zha86,AncBlu96,AncBlu97}, 
the identity \eqref{frechetid} yields a conserved current 
when $f=G^a$, $g=P^\alpha$, $h=Q_a$, 
where $P^\alpha$ is the characteristic of an infinitesimal symmetry \eqref{symmchar}
and $Q_a$ is an adjoint-symmetry, 
which satisfy $(\delta_P G)^a|_\E=0$ and $(\delta_Q^* G)_\alpha|_\E=0$. 
Note, by \lemref{hadamard}, it follows that 
\begin{subequations}\label{deteqoffE}
\begin{align}
& (\delta_P G)^a=R_P(G)^a
\label{symmdeteqoffE}
\\
& (\delta_Q^* G)_\alpha=R_Q(G)_\alpha
\label{adjsymmdeteqoffE}
\end{align}
\end{subequations}
each hold identically, 
where $R_Q$ is the linear differential operator \eqref{RQop},
and $R_P$ is a similar linear differential operator
\begin{equation}\label{RPop}
(R_P)^a =
R_P^{(0)}{}^{a}_{b} + R_P^{(1)}{}^{a}_{b} D_t + R_P^{(1)}{}^{ai}_{b} D_i 
+ R_P^{(2)}{}^{a}_{b} D_t{}^2 + R_P^{(2)}{}^{ai}_{b} D_tD_i + R_P^{(2)}{}^{aij}_{b} D_iD_j +\cdots . 
\end{equation}
In both operators $R_P$ and $R_Q$, 
the coefficients are non-singular on $\E$ 
as the PDE system $G^a=0$ is assumed to be normal.

We can now state the first main result 
for PDE systems with any number of variables,
which extends the results derived in recent work \cite{Anc16a}
on single PDEs with two independent variables. 

\begin{thm}\label{symmactionTXQ}
For a given normal PDE system \eqref{pde},
let $\Phi=(T,X^i)$ be a conserved current and $Q_a$ be its multiplier,
and let $\hat\X=P^\alpha\parderop{u^\alpha}$ be an infinitesimal symmetry. 
Then $\Psi_G(P,Q)=(\Psi_G^t(P,Q),\Psi_G^i(P,Q))$ defines a conserved current 
which is locally equivalent to the conserved current obtained by the symmetry action 
$\Phi_{\hat\X} = (T_{\hat\X},X^i_{\hat\X})= \pr\hat\X(\Phi)$ on $\Phi$. 
In explicit form, 
\begin{equation}\label{symmactionT}
\begin{aligned}
T_{\hat\X}= \delta_P T & = \Psi_G^t(P,Q) 
\\
& = P^\alpha\Big( 
Q_a \parder{G^a}{u^\alpha_t} -D_t\Big(Q_a \parder{G^a}{u^\alpha_{tt}}\Big) -D_i\Big(Q_a \parder{G^a}{u^\alpha_{tx\up{i}}}\Big) +\cdots 
\Big)
\\&\qquad
+ D_t P^\alpha\Big( 
Q_a\parder{G^a}{u^\alpha_{tt}} -D_t\Big(Q_a\parder{G^a}{u^\alpha_{ttt}}\Big) -D_i\Big(Q_a \parder{G^a}{u^\alpha_{ttx\up{i}}}\Big) +\cdots 
\Big)
\\&\qquad
+ D_t{}^2 P^\alpha\Big( 
Q_a\parder{G^a}{u^\alpha_{ttt}} -D_t\Big(Q_a\parder{G^a}{u_{tttt}}\Big) -D_i\Big(Q_a\parder{G^a}{u^\alpha_{tttx\up{i}}}\Big) +\cdots 
\Big)
+\cdots 
\end{aligned}
\end{equation}
modulo trivial terms $D_i\Theta^i$,
and 
\begin{equation}\label{symmactionX}
\begin{aligned}
X^i_{\hat\X} = (\delta_P X)^i & = \Psi_G^i(P,Q) 
\\
& = P^\alpha\Big( 
Q_a \parder{G^a}{u^\alpha_{x\up{i}}} -D_j\Big(Q_a\parder{G^a}{u^\alpha_{x\up{i}x\up{j}}}\Big) +D_jD_k\Big(Q_a\parder{G^a}{u^\alpha_{x\up{i}x\up{j}x\up{k}}}\Big) +\cdots 
\Big)
\\&\qquad
+ D_t P^\alpha\Big( 
Q_a \parder{G^a}{u^\alpha_{tx\up{i}}} -D_j\Big(Q_a\parder{G^a}{u^\alpha_{tx\up{i}x\up{j}}}\Big) +D_jD_k\Big(Q_a\parder{G^a}{u^\alpha_{tx\up{i}x\up{j}x\up{k}}}\Big) +\cdots 
\Big)
\\&\qquad
+ D_j P^\alpha\Big( 
Q_a\parder{G^a}{u^\alpha_{x\up{i}x\up{j}}} -D_k\Big(Q_a\parder{G^a}{u^\alpha_{x\up{i}x\up{j}x\up{k}}}\Big) +D_kD_l\Big(Q_a\parder{G^a}{u^\alpha_{x\up{i}x\up{j}x\up{k}x\up{l}}}\Big) +\cdots 
\Big)
\\&\qquad
+\cdots 
\end{aligned}
\end{equation}
modulo trivial terms $-D_t\Theta^i+D_j\Gamma^{ij}$.
The multiplier of this conserved current is given by 
\begin{equation}\label{symmactionQ}
\begin{aligned}
Q_a^{\hat\X} & = R_P^*(Q)_a -R_Q^*(P)_a
\\
& = \Big( 
R_P^{(0)}{}^{b}_{a}Q_b -D_t(R_P^{(1)}{}^{b}_{a} Q_b) - D_i(R_P^{(1)}{}^{bi}_{a} Q_b)
+ D_t{}^2(R_P^{(2)}{}^{b}_{a}Q_b) 
\\&\qquad
+D_tD_i(R_P^{(2)}{}^{bi}_{a} Q_b) + D_iD_j(R_P^{(2)}{}^{bij}_{a} Q_b) 
+\cdots \Big)
\\&\qquad
-\Big( 
R_Q^{(0)}{}_{a\alpha}P^\alpha  - D_t(R_Q^{(1)}{}_{a\alpha}P^\alpha) - D_i(R_Q^{(1)}{}^i_{a\alpha} P^\alpha) 
+ D_t{}^2(R_Q^{(2)}{}_{a\alpha} P^\alpha) 
\\&\qquad\qquad
+ D_tD_i(R_Q^{(2)}{}^i_{a\alpha} P^\alpha) + D_iD_j(R_Q^{(2)}{}^{ij}_{a\alpha} P^\alpha) 
+\cdots \Big) . 
\end{aligned}
\end{equation}
Hence, 
the conserved currents $\Psi_G(P,Q)$ and $\Phi_{\hat\X}$ are trivial 
iff $Q_a^{\hat\X} = 0$, $a=1,\ldots,M$, vanishes identically. 
\end{thm}

This result has a straightforward proof by comparing the multipliers 
for $\Psi_f(P,Q)$ and $\Phi_{\hat\X}$. 
Consider the local conservation law determined by the multiplier $Q_a$.
The symmetry $\hat\X$ applied to the characteristic equation \eqref{chareqn} of this conservation law
yields
\begin{equation}\label{X_on_chareqn}
\pr\hat\X(D_t\tilde T + D_i\tilde X^i) = \pr\hat\X(Q_aG^a)
= \delta_P(Q_aG^a) = (\delta_P Q)_a G^a + Q_a(\delta_P G^a) .
\end{equation}
The last term in this equation can be expressed as
\begin{equation}\label{X_on_chareqn_term2}
Q_a(\delta_P G^a) = Q_a R_P(G)^a = G^a R_P^*(Q)_a + D_t \Upsilon^t(Q,G;P) + D_i \Upsilon^i(Q,G;P)
\end{equation}
using the symmetry identity \eqref{symmdeteqoffE} combined with integration by parts,
where $\Upsilon^t(Q,G;P)|_\E =0$  and $\Upsilon^i(Q,G;P)|_\E =0$.
Next,
the second-last term in equation \eqref{X_on_chareqn} can be expressed as
\begin{equation}\label{X_on_Q_term1}
(\delta_P Q)_a G^a  = P^\alpha (\delta^*_G Q)_\alpha + D_t \Psi_Q^t(P,G) + D_i\Psi_Q^i(P,G)
\end{equation}
through the Frechet derivative identity \eqref{frechetid}.
Then, through the multiplier determining equation \eqref{multdeteq}
and the adjoint-symmetry identity \eqref{adjsymmdeteqoffE}, 
the first term in equation \eqref{X_on_Q_term1} becomes 
\begin{equation}
P^\alpha (\delta^*_G Q)_\alpha = -P^\alpha (\delta^*_{Q}G)_\alpha 
= -P^\alpha R_{Q}(G)_\alpha  . 
\end{equation}
This yields
\begin{equation}
(\delta_P Q)_a G^a  = -P^\alpha R_{Q}(G)_\alpha + D_t \Psi_Q^t(P,G) + D_i\Psi_Q^i(P,G) . 
\end{equation}
Integration by parts on the first term in this equation gives 
\begin{equation}\label{X_on_chareqn_term1}
(\delta_P Q)_a G^a  = -G^a R^*_{Q}(P)_a + D_t(\Psi_Q^t(P,G) -\Upsilon^t(P,G;Q)) + D_i(\Psi_Q^i(P,G) -\Upsilon^i(P,G;Q)) 
\end{equation}
where $\Upsilon^t(P,G;Q)|_\E =0$  and $\Upsilon^i(P,G;Q)|_\E =0$.
Substitution of expressions \eqref{X_on_chareqn_term1} and \eqref{X_on_chareqn_term2}
into equation \eqref{X_on_chareqn} now yields
\begin{equation}\label{X_on_chareqn_terms}
\pr\hat\X(D_t\tilde T + D_i\tilde X^i) 
= ( R_P^*(Q)_a  - R^*_{Q}(P)_a ) G^a + D_t \tilde\Upsilon^t + D_i \tilde\Upsilon^i
\end{equation}
where
\begin{equation}
\tilde\Upsilon^t = \Psi_Q^t(P,G) +\Upsilon^t(Q,G;P)-\Upsilon^t(P,G;Q), 
\quad
\tilde\Upsilon^i = \Psi_Q^i(P,G) +\Upsilon^i(Q,G;P)-\Upsilon^i(P,G;Q) 
\end{equation}
is a trivial conserved current since 
$\tilde\Upsilon^t|_\E = 0$ and $\tilde\Upsilon^i|_\E = 0$. 
Finally, since $\pr\hat\X$ commutes with total derivatives \cite{Olv,Anc-review},
equation \eqref{X_on_chareqn_terms} becomes 
\begin{equation}\label{X_on_conslaw}
D_t (T_{\hat\X} -\hat\Upsilon^t) +D_i( X^i_{\hat\X} -\hat\Upsilon^i) 
= Q_a^{\hat\X} G^a
\end{equation}
with 
\begin{equation}
\hat\Upsilon^t =\tilde\Upsilon^t + T-\tilde T,
\quad
\hat\Upsilon^i =\tilde\Upsilon^i + X^i -\tilde X^i, 
\end{equation}
where $Q_a^{\hat\X}$ is given by expression \eqref{symmactionQ}. 
Since $\hat\Upsilon^t|_\E =0$ and $\hat\Upsilon^i|_\E =0$, 
equation \eqref{X_on_conslaw} is the characteristic equation of the conserved current \eqref{symmactionT}--\eqref{symmactionX}.
This completes the proof.

\thmref{symmactionTXQ} can be used to provide a direct characterization 
for when a conservation law is invariant, or more generally is homogeneous, 
under the action of a symmetry, 
as defined in \Ref{Anc16a}. 
For a given infinitesimal symmetry \eqref{symmchar}, 
a conservation law \eqref{conslaw} is {\em symmetry-invariant} iff 
the symmetry action on the corresponding conserved current $\Phi=(T,X^i)$ 
yields a trivial current, 
\begin{equation}\label{invconslaw}
\Phi_{\X}|_\E=(\pr\hat\X(T)|_\E,\pr\hat\X(X^i)|_\E) =(D_i\Theta^i,-D_t\Theta^i+D_j\Gamma^{ij})|_\E .
\end{equation}
(This generalizes the notion introduced in \Ref{KarMah00} 
for singling out conserved currents that are strictly unchanged 
under the action of a symmetry.)
A natural extension is to allow a conserved current to be locally equivalent to 
a multiple of itself under the symmetry action, 
\begin{equation}\label{homoconslaw}
\Phi_{\X}|_\E-\lambda \Phi|_\E 
=((\pr\hat\X(T) -\lambda T)|_\E,(\pr\hat\X(X^i) -\lambda X^i)|_\E) + (D_i\Theta^i,-D_t\Theta^i+D_j\Gamma^{ij})|_\E
\end{equation}
where $\lambda$ is a non-zero constant.
This corresponds to a a conservation law \eqref{conslaw} being 
{\em symmetry-homogeneous}
under the action of an infinitesimal symmetry \eqref{symm}. 

The condition of symmetry-invariance and symmetry-homogeneity 
for conservation laws has a simple formulation in terms of multipliers, 
which follows immediately from \thmref{symmactionTXQ}. 

\begin{thm}\label{symmcond}
A conservation law \eqref{conslaw} is homogeneous \eqref{homoconslaw}
under the action of an infinitesimal symmetry \eqref{symm} 
iff its multiplier \eqref{Q} satisfies the condition 
\begin{equation}\label{invcond}
R_P^*(Q)_a -R_Q^*(P)_a=\lambda Q_a
\end{equation}
for some constant $\lambda$. 
The conservation law is invariant \eqref{invconslaw}
iff $\lambda=0$. 
If a PDE system is an Euler-Lagrange system \eqref{ELpde}, 
then every conservation law is invariant under the variational symmetry 
corresponding to its multiplier. 
\end{thm}

\begin{cor}
(i) 
Under the action of an infinitesimal symmetry \eqref{symm},
a conserved quantity \eqref{C} on a spatial domain $\Omega\subseteq\Rnum^n$ 
is unchanged modulo an arbitrary boundary term,
$\hat\X C[u] = \int_{\p\Omega} \Theta^i dA_i$, 
iff the corresponding conservation law \eqref{conslaw} is symmetry invariant. 
(ii) 
A conserved quantity \eqref{C} is mapped into itself 
(modulo an arbitrary boundary term) under a symmetry 
iff the corresponding conservation law \eqref{conslaw} is symmetry homogeneous. 
In particular, 
under the action of a symmetry transformation $\exp(\epsilon\hat\X)$
with parameter $\epsilon$, 
a symmetry-homogeneous conserved quantity $C[u]$ is mapped into 
$\exp(\epsilon\lambda)C(u)$ whenever all boundary terms vanish. 
\end{cor}

One very useful application of these results is that 
the formula \eqref{symmactionT}--\eqref{symmactionX} can be used to 
construct the conserved current determined by a given multiplier 
in the important case when a PDE system admits a scaling symmetry. 
This was first developed and applied in \Ref{Anc03}. 
Here we summarize the main result which will be used in the examples 
in the next section. (See \Ref{Anc16a,Anc03} for a proof.)

\begin{prop}\label{PQformula}
Suppose a normal PDE system \eqref{pde} possesses a scaling symmetry
\begin{equation}
t\rightarrow \lambda^p t,
\quad
x^i\rightarrow \lambda^{q^{(i)}} x^i,
\quad
u^\alpha\rightarrow \lambda^{r^{(\alpha)}}u^\alpha
\end{equation}
where $p,q^{(i)},r^{(\alpha)}$ are constants.
Let $Q_a$ be the multiplier for a conservation law in which the components of
the conserved current $\Phi=(T,X^i)$ are scaling homogeneous,
$(T,X^i) \rightarrow (\lambda^k T,\lambda^{k^{(i)}}X^i)$.
Then, in terms of the characteristic functions 
$P^\alpha=r^{(\alpha)}u^\alpha-(ptu^\alpha_t+q^{(i)}x^i u_{x\up{i}})$ 
of the scaling symmetry, 
the conserved current $\Psi_G(P,Q)$ is equivalent to a multiple 
$w=k+\sum_{i=1}^{n}q^{(i)}$ of the conserved current $\Phi$. 
This multiple $w$ is equal to the scaling weight of the conserved quantity 
$C[u]=\int_{\Omega} T d^nx$.
Hence, 
in the case when the conserved quantity is homogeneous under the scaling symmetry, 
so that $w\neq0$, 
the components of the conserved current are given by 
$\Phi=(1/w)\Psi_G(P,Q)$ up to equivalence. 
\end{prop}

\section{Examples}
\label{examples}

We will now consider several different examples of PDEs and PDE systems
arising in a variety of physical applications,
including water waves, fluid flow, and gas dynamics. 
For each example, 
we first show how to set up and apply the multiplier method 
to obtain all low-order conservation laws, 
and next we examine the symmetry properties of these conservation laws.
All calculations have been carried out in Maple.

\subsection{Nonlinear hyperbolic equation}
\label{gMT-ex}

Our first example is the nonlinear PDE
\begin{equation}\label{gMT-eqn}
u_{tx} + u_x + ku^p = 0,
\quad
p>1, k\neq 0 . 
\end{equation}
This is a hyperbolic equation which arises, for $p=2$, 
from gas dynamics when a generating function is formulated 
for the moments of the gas velocity distribution function 
exhibiting a Maxwellian tail \cite{KroWu}. 
We will refer to the PDE \eqref{gMT-eqn} 
as the generalized Maxwellian tails (gMT) equation. 

While the gMT equation is not an Euler-Lagrange equation as it stands, 
it does come from a Lagrangian
$L = e^{2t}( \tfrac{1}{2} u_t u_x -\tfrac{k}{p+1} u^{p+1} )$ 
through a variational integrating factor $e^{2t}$, 
as shown by $E_u(L)= -e^{2t}G$ where $G=u_{tx} + u_x + ku^p$. 
In particular, under a related point transformation, 
the gMT equation can be transformed into a Klein-Gordon equation 
which is a normal PDE. 
However, we will show how to obtain the conservation laws of the gMT equation directly, 
without the need to use any transformations. 

The determining equation \eqref{symmdeteq} 
for infinitesimal symmetries $\X=P\p_u$ 
of the gMT equation \eqref{gMT-eqn} is given by 
\begin{equation}
(D_tD_xP + D_xP + kpu^{p-1}P)|_\E=0 . 
\end{equation}
A straightforward computation of point and contact symmetries 
yields the characteristic functions 
\begin{equation}
P_1 = -u_t, 
\quad
P_2 = -u_x, 
\quad
P_3 = e^{(p-1)t}(u_t + u), 
\quad
P_4 = (p-1)xu_x + u, 
\end{equation}
from which we obtain
\begin{equation}\label{gMT-RP}
R_{P_1} = -D_t,
\quad
R_{P_2} = -D_x,
\quad
R_{P_3} = e^{(p-1)t}(D_t +p),
\quad
R_{P_4} = (p-1)xD_x + p
\end{equation}
as given by equation \eqref{deteqoffE}. 
These symmetries consist of separate translations in $t$ and $x$, 
a time-dependent dilation in $u$ combined with a time-dependent shift in $t$,
and a scaling in $u$ and $x$, 
all of which are point symmetries \cite{EulLeaMahSte}
\begin{equation}\label{gMT-pointsymms}
\X_1 = \p_t, 
\quad
\X_2 = \p_x, 
\quad
\X_3 = e^{(p-1)t}(-\p_t + u\p_u), 
\quad
\X_4 = (1-p)x\p_x +  u\p_u , 
\end{equation}
with the respective group actions 
\begin{align}
& t\rightarrow t+\epsilon,
\label{gMT-pt1}\\
& x\rightarrow x+\epsilon,
\label{gMT-pt2}\\
& t\rightarrow \tfrac{1}{1-p}\ln((p-1)\epsilon+e^{(1-p)t}),
\quad
u\rightarrow ((p-1)\epsilon+e^{(1-p)t})^{p-1}u, 
\label{gMT-pt3}\\
& x\rightarrow e^{(p-1)\epsilon}x,
\quad
u\rightarrow e^{\epsilon}u , 
\label{gMT-pt4}
\end{align}
in terms of a parameter $\epsilon$. 

Since the gMT equation \eqref{gMT-eqn} has a Lagrangian formulation, 
all conservation laws of this equation 
arise from multipliers given by the relation 
\begin{equation}\label{gMT-PQrel}
Q= e^{2t}P
\end{equation}
where $P$ is the characteristic of a variational symmetry $\hat\X=P\p_u$ 
of the gMT equation. 
Variational contact symmetries and variational point symmetries
thereby correspond to multipliers with the form 
\begin{equation}\label{gMT-1stordQ}
Q(t,x,u,u_t,u_x) . 
\end{equation}
Note that $u_t,u_x$ are the only derivatives of $u$ that can be differentiated to obtain the leading derivative $u_{tx}$ in the gMT equation,
and hence multipliers having the form \eqref{gMT-1stordQ} generate 
all low-order conservation laws for the gMT equation. 
The determining system for these multipliers \eqref{gMT-PQrel}--\eqref{gMT-1stordQ} 
consists of the adjoint-symmetry determining equation \eqref{adjsymmdeteq} 
which is given by 
\begin{equation}
(D_tD_xQ - D_xQ + kpu^{p-1}Q)|_\E=0, 
\end{equation}
and the Helmholtz equations \eqref{helmholtzeq} 
which can be shown to reduce to the equation 
\begin{equation}
Q_u -Q_{u_t}=0 . 
\end{equation}
Hence, 
a point or contact symmetry of the gMT equation \eqref{gMT-eqn} is variational iff 
its characteristic satisfies the condition 
\begin{equation}\label{gMT-varsymmcond}
P_u -P_{u_t}=0 . 
\end{equation}
From \thmref{adjnoether}, 
this condition is equivalent to invariance of the Lagrangian, 
$\pr\hat\X(L) = D_t A + D_x B$ for some differential functions 
$A(t,x,u,u_t,u_x)$ and $B(t,x,u,u_t,u_x)$. 

When the variational symmetry condition \eqref{gMT-varsymmcond} is applied to 
a linear combination of the four point symmetries \eqref{gMT-pointsymms}, 
it shows that $\X_1 -\X_4$, $\X_2$, $\X_3$ 
generate all variational point symmetries. 
These three variational symmetries correspond to the respective multipliers
\begin{equation}\label{gMT-Q}
Q_1 = -e^{2t}(u_t + (p-1)xu_x + u),
\quad
Q_2 = -e^{2t}u_x, 
\quad
Q_3 = e^{(p+1)t}(u_t + u), 
\end{equation}
from which we obtain
\begin{equation}\label{gMT-RQ}
R_{Q_1} = -e^{2t}((p-1)x D_x+D_t+p),
\quad
R_{Q_2} = -e^{2t}D_x, 
\quad
R_{Q_3} = e^{(p+1)t}(D_t + p)
\end{equation}
as given by equation \eqref{deteqoffE}. 

Each multiplier \eqref{gMT-Q} satisfies the characteristic equation
\begin{equation}\label{gMT-conslaw}
D_t T + D_x X = QG,
\quad
Q=T_{u_x} + X_{u_t},
\quad
G=u_{tx} + u_x + ku^p
\end{equation}
where the conserved current $\Phi=(T(t,x,u,u_t,u_x),X(t,x,u,u_t,u_x))$ 
can be obtained directly by 
either integration of the characteristic equation \eqref{gMT-conslaw}
or use of the scaling formula $\Psi_G((p-1)xu_x + u,Q)$ 
from \propref{PQformula}. 
In particular, the components of the scaling formula are given by 
\begin{equation}
\Psi_G^t = -((p-1)xu_x + u)D_xQ,
\quad
\Psi_G^x = (u_t-(p-1)kxu^p + u)Q . 
\end{equation}
We obtain, modulo the addition of a trivial current, 
\begin{align}
& 
T_1= e^{2t}( \tfrac{p-1}{2}xu_x^2 + \tfrac{k}{p+1} u^{p+1}),
\quad
X_1 = e^{2t}( \tfrac{1}{2}(u+u_t)^2 + \tfrac{k(p-1)}{p+1}x u^{p+1})
\label{gMT-TX-1}\\
&
T_2 = \tfrac{1}{2} e^{2t}u_x^2,
\quad
X_2 = \tfrac{k}{p+1} e^{2t} u^{p+1}
\label{gMT-TX-2}\\
&
T_3 = \tfrac{1}{p+1}e^{(p+1)t}u^{p+1},
\quad
X_3 = \tfrac{1}{2} e^{(p+1)t} (u+u_t)^2 
\label{gMT-TX-3}
\end{align}
These conserved currents represent conservation of 
three energy-momentum quantities $\int_\Omega T_i dx$ 
for the gMT equation \eqref{gMT-eqn} 
on any spatial domain $\Omega\subseteq\Rnum$.

We will now study the symmetry properties of the conservation laws 
\eqref{gMT-TX-1}--\eqref{gMT-TX-3}. 
Consider the vector space of conserved currents
\begin{equation}\label{gMT-gencurrent}
T = a_1 T_1 +a_2 T_2 +a_3 T_3,
\quad
a_i=\const , 
\end{equation}
and the algebra of point symmetries
\begin{equation}\label{gMT-gensymm}
\X = c_1\X_1+c_2 \X_2+c_3 \X_3+c_4 \X_4,
\quad
c_j=\const . 
\end{equation}
A conservation law $(D_tT+D_xX)|_\E=0$ is homogeneous under the symmetry $\X$ 
iff condition \eqref{invcond} is satisfied, 
where the characteristic function of the symmetry generator is given by 
$P= c_1P_1+c_2P_2+c_3 P_3+c_4P_4$,
and the multiplier for the conservation law is given by 
$Q= a_1Q_1+a_2Q_2+a_3 Q_3$. 
By using equations \eqref{gMT-RP} and \eqref{gMT-RQ}, 
we find that the condition \eqref{invcond} splits with respect to $t,x,u,u_t,u_x$ 
into a system of bilinear equations on $c_j$ and $a_i$:
\begin{subequations}\label{gMT-aceqs}
\begin{align}
& a_1(\lambda-2c_1-2c_4) =0
\\
& ((p+1)c_4+2c_1-\lambda)a_2+(p-1)a_1c_2 =0
\\
& ((p+1)c_1+2c_4-\lambda)a_3-(p-1)a_1c_3 =0
\end{align}
\end{subequations}
The solutions for $c_j$ in terms of $a_i$ 
determine the symmetry-homogeneity properties of 
the conserved current $\Phi=(T,X)$ modulo trivial currents. 
Note that any solution can be scaled by a constant factor, 
$c_j\rightarrow \gamma c_j$ with $\gamma\neq 0$. 
By considering the subspaces generated by $\{a_i\}$,
solving the system \eqref{gMT-aceqs} in each case, 
and merging the solutions, 
we obtain the conditions
\begin{subequations}
\begin{align}
& 
a_1{}^2+a_2{}^2+a_3{}^2\neq 0;
\quad
a_3 c_1- a_1 c_3 =0,
\quad
a_1c_2 + a_2c_4 =0, 
\quad
\lambda = 2(c_1+c_4)
\\
& 
a_2\neq 0,
\quad
a_1 = a_3 = 0;
\quad
\lambda=(p+1)c_4+2 c_1
\\
& 
a_3\neq 0,
\quad
a_1 = a_2 = 0;
\quad
\lambda=(p+1)c_1+2 c_4
\end{align}
\end{subequations}
Hence, we conclude the following.

(1) 
The symmetry properties of the vector space 
$a_1\Phi_1 +a_2\Phi_2+a_3\Phi_3+\Phi_\triv$ for arbitrary $a_i$ consist of:
(i) invariance under $a_1(\X_1-\X_4)+a_2\X_2 +a_3\X_3$;
(ii) homogeneity under $a_1\X_1+a_3\X_3$ and $a_1\X_4-a_2\X_2$, with $\lambda=2a_1$.

(2) 
The only additional symmetry properties of the vector space 
$a_1\Phi_1 +a_2\Phi_2+a_3\Phi_3+\Phi_\triv$ consist of:
(i) invariance of the subspace $a_2\Phi_2+\Phi_\triv$ 
under $\X_3$ and $2\X_4-(p+1)\X_1$;
(ii) invariance of the subspace $a_3\Phi_3+\Phi_\triv$ 
under $\X_2$ and $2\X_1-(p+1)\X_4$;
(iii) homogeneity of the subspace $a_2\Phi_2+\Phi_\triv$ 
under $\X_4$ with $\lambda =p+1$, and $\X_1$ with $\lambda=2$;
(iv) homogeneity of the subspace $a_3\Phi_3+\Phi_\triv$ 
under $\X_1$ with $\lambda =p+1$, and $\X_4$ with $\lambda=2$. 

From these properties, 
it is simple to work out the symmetries for which each energy-momentum quantity 
$\int_\Omega T_i dx$ is invariant (modulo an endpoint term $\Theta|_{\p\Omega}$).
In particular, 
$\int_\Omega T_1 dx$ is invariant under $\X_1-\X_4$;
$\int_\Omega T_2 dx$ is invariant under $\X_2$, $\X_3$, $\X_4-\tfrac{1}{2}(p+1)\X_1$;
and $\int_\Omega T_3 dx$ is invariant under $\X_2$, $\X_3$, $\X_1-\tfrac{1}{2}(p+1)\X_4$. 
More generally, 
with boundary conditions such that all endpoint terms vanish, 
$\int_\Omega T_1 dx$ is mapped into $e^{2\epsilon}\int_\Omega T_1 dx$ 
under time-translations \eqref{gMT-pt1} and scalings \eqref{gMT-pt4};
$\int_\Omega T_2 dx$ is mapped 
into $e^{2\epsilon}\int_\Omega T_2 dx$ under time-translations \eqref{gMT-pt1},  
and into $e^{(p+1)\epsilon}\int_\Omega T_2 dx$ under scalings \eqref{gMT-pt4};
$\int_\Omega T_3 dx$ is mapped 
into $e^{(p+1)\epsilon}\int_\Omega T_3 dx$ under time-translations \eqref{gMT-pt1}, 
and into $e^{2\epsilon}\int_\Omega T_3 dx$ under scalings \eqref{gMT-pt4}. 

It is interesting to note that every conserved quantity 
$\int_\Omega a_1T_1 + a_2T_2 +a_3T_3 dx$ is invariant 
under the variational symmetry $\hat\X=e^{-2t}Q\p_u$ that corresponds to 
its multiplier $Q=a_1Q_1 + a_2Q_2 +a_3Q_3$ 
through the relation \eqref{gMT-PQrel}. 
This relation also shows that $R_Q=e^{2t}R_P$, 
from which we see that the symmetry-invariance condition $R_P^*(Q)-R_Q^*(P)=0$
holds identically when $Q= e^{2t}P$. 
Consequently, 
the symmetry-invariance property stated in \thmref{symmcond} 
for conservation laws of Euler-Lagrange PDEs 
extends to the case of PDEs that acquire the form of Euler-Lagrange equations
when a variational integrating factor is introduced. 

As a final result, 
we observe that all three energy-momentum conservation laws \eqref{gMT-TX-1}--\eqref{gMT-TX-3} 
are homogeneous under the non-variational symmetries $\X_1$ and $\X_4$,
while under the variational symmetry $\X_1-\X_4$, 
the two conservation laws \eqref{gMT-TX-2} and \eqref{gMT-TX-3} 
are homogeneous and the conservation law \eqref{gMT-TX-1} is invariant. 
Thus, in contrast to Noether’s theorem, 
the symmetry homogeneity condition for multipliers 
can yield a conservation law from a non-variational symmetry
and also can yield more than one conservation law from a single variational symmetry.

\subsection{Dispersive nonlinear wave equation}
\label{gKdV-ex}

Our next example is the generalized Korteveg-de Vries (gKdV) equation
\begin{equation}\label{gKdV-eqn}
u_t + u_{xxx} + ku^p u_x=0,
\quad
p>0, k\neq 0 . 
\end{equation}
This PDE is a dispersive nonlinear wave equation, 
which reduces to the KdV equation when $p=1$ 
and the modified KdV equation when $p=2$. 
If a potential $v$ is introduced by $u=v_x$, 
then the Lagrangian 
$L = -\frac{1}{2}v_t v_x +\frac{1}{2}v_{xx}^2+ \frac{k}{(p+1)(p+2)}v_x^{p+2}$
yields $E_v(L) = v_{tx} + v_{xxxx} + kv_x^p v_{xx}=G$, 
where $G=u_t + u_{xxx} + ku^p u_x$. 
Note this is a normal PDE, 
as it has the leading derivative $u_{t}$ or $u_{xxx}$. 

The determining equation \eqref{symmdeteq} 
for infinitesimal symmetries $\X=P\p_u$ 
of the gKdV equation \eqref{gKdV-eqn} is given by 
\begin{equation}\label{gKdV-symmeq}
(D_tP +D_x{}^3P +ku^pD_x P +kpu^{p-1}P)|_\E=0 . 
\end{equation}
It is well known that the gKdV equation has no contact symmetries 
and that all of its point symmetries are generated by 
a time translation, a space translation, and a scaling, 
when $p\neq 0$, 
plus a Galilean boost, when $p=1$. 
For these symmetries
\begin{gather}
\X_1 = \p_t, 
\quad
\X_2 = \p_x, 
\quad
\X_3 = 3t\p_t + x\p_x -(2/p)u\p_u, 
\label{gkdv-pointsymms}\\
\X_4 = kt\p_x +\p_u, 
\quad
p=1 , 
\label{gkdv-pointsymms-p=1}
\end{gather}
the characteristic functions are given by 
\begin{equation}
\begin{gathered}
P_1 = -u_t, 
\quad
P_2 = -u_x, 
\quad
P_3 = -(3tu_t +xu_x + (2/p)u), 
\\
P_4 = 1-ktu_x,
\quad
p=1, 
\end{gathered}
\end{equation}
from which we obtain
\begin{equation}\label{gKdV-RP}
R_{P_1} = -D_t,
\quad
R_{P_2} = -D_x,
\quad
R_{P_3} = -(3tD_t +xD_x + (3+2/p)),
\quad
R_{P_4} = -ktD_x
\end{equation}
as given by equation \eqref{deteqoffE}. 
Their respective group actions are given by 
\begin{align}
& t\rightarrow t+\epsilon,
\label{gKdV-pt1}\\
& x\rightarrow x+\epsilon,
\label{gKdV-pt2}\\
& t\rightarrow e^{3\epsilon}t,
\quad
x\rightarrow e^{\epsilon}x,
\quad
u\rightarrow e^{-(2/p)\epsilon}u, 
\label{gKdV-pt3}\\
& x\rightarrow x+\epsilon kt,
\quad
u\rightarrow u+\epsilon, 
\label{gKdV-pt4}
\end{align}
in terms of a parameter $\epsilon$. 

The characteristic equation for conservation laws of the gKdV equation \eqref{gKdV-eqn} is given by 
\begin{equation}\label{gKdV-conslaw}
D_t T + D_x X = QG,
\quad
Q = E_u(T)
\end{equation}
where $T$ is a function of $t,x,u$, and $x$-derivatives of $u$,
with all $t$-derivatives of $u$ being eliminated from $T$ 
through the gKdV equation $u_t=-(u_{xxx} + ku^p u_x)$. 
Thus, multipliers $Q$ will be functions only of $t,x,u$, 
and $x$-derivatives of $u$.
The determining system \eqref{adjsymmdeteq} and \eqref{helmholtzeq} 
for multipliers then consists of 
the adjoint-symmetry equation 
\begin{equation}\label{gKdV-adjsymmeq}
-(D_tQ +D_x{}^3Q +ku^pD_x Q)|_\E=0
\end{equation}
and the Helmholtz equations 
\begin{equation}\label{gKdV-helmhotzeq}
Q_u = E_u(Q), 
\quad
Q_{u_x} = -E_u^{(1,x)}(Q), 
\quad
Q_{u_{xx}} = E_u^{(2,xx)}(Q), 
\quad
\ldots .
\end{equation}
Since the gKdV equation \eqref{gKdV-eqn} is an Euler-Lagrange equation 
in terms of the potential $v$, 
all multipliers are the same as characteristics of variational symmetries 
$\hat\X = Q\p_v$ acting on $v$. 
When prolonged to $u=v_x$, these symmetries have the characteristic form 
\begin{equation}\label{gKdV-PQrel}
P= D_x Q
\end{equation}
corresponding to $\hat\X= D_xQ\p_u$, 
where $P$ contains no $t$-derivatives of $u$. 

The characteristic functions of the point symmetries \eqref{gkdv-pointsymms}--\eqref{gkdv-pointsymms-p=1}
of the gKdV equation \eqref{gKdV-eqn}
are given by, after elimination of $t$-derivatives of $u$, 
\begin{equation}\label{gKdv-pointP}
\begin{gathered}
P_1 = u_{xxx} + ku^p u_x,
\quad
P_2 = -u_x, 
\quad
P_3 = 3tu_{xxx} + (3tku^p -x)u_x -(2/p)u,
\\
P_4 = 1-ktu_x,
\quad
p=1 , 
\end{gathered}
\end{equation}
which are each of third order. 
It is now straightforward to derive the conditions 
for a point symmetry of the gKdV equation \eqref{gKdV-eqn} to yield
a conservation law multiplier. 
First, the correspondence relation \eqref{gKdV-PQrel} holds iff 
\begin{equation}\label{gKdV-Pcond1}
0=E_u(P) = P_u -D_x P_{u_x} + D_x{}^2 P_{u_{xx}} - D_x{}^3 P_{u_{xxx}} . 
\end{equation}
Second, from this relation \eqref{gKdV-PQrel}, 
the symmetry determining equation \eqref{gKdV-symmeq} becomes 
$D_x(D_tQ +D_x{}^3Q +ku^pD_x Q)|_\E=0$
where $Q=D_x^{-1}P$ is of second order. 
This equation implies $(D_tQ +D_x{}^3Q +ku^pD_x Q)|_\E=f(t)$,
but since $P$ is a homogeneous function (\ie/ $P|_{u=0}=0$), 
then $Q$ can be assumed to be homogeneous,
yielding $f(t)=0$. 
Hence $Q$ thereby satisfies the adjoint-symmetry equation \eqref{gKdV-adjsymmeq}. 
Next, since $Q$ is of second order, 
the Helmholtz equations \eqref{gKdV-helmhotzeq} reduce to the equation 
\begin{equation}\label{gKdV-Qcond}
Q_{u_x} -D_x Q_{u_{xx}}=0 . 
\end{equation}
Both this equation \eqref{gKdV-Qcond} and equation \eqref{gKdV-PQrel}
now can be split with respect to all derivatives of $u$ 
higher than second-order. 
This splitting yields $P_{u_{xxx}} = Q_{u_{xx}}$ and $P_{u_{xx}} = 2Q_{u_x}$,
from which it can be shown that the Helmholtz equation \eqref{gKdV-Qcond}
is equivalent to the condition
\begin{equation}\label{gKdV-Pcond2}
P_{u_{xx}} -2D_x P_{u_{xxx}}=0 . 
\end{equation}
Therefore, the conditions \eqref{gKdV-Pcond2} and \eqref{gKdV-Pcond1}
are necessary and sufficient for a point symmetry to yield a multiplier
through the correspondence relation \eqref{gKdV-PQrel}. 
The resulting multipliers 
\begin{equation}\label{gKdV-2ndordQ}
Q(t,x,u,u_x,u_{xx}) 
\end{equation}
will correspond to variational point symmetries 
$\hat\X = Q\p_v$ for the gKdV equation expressed in Lagrangian form 
using the potential $u=v_x$. 
Each multiplier determines a conserved current 
\begin{equation}\label{gKdV-TX}
\Phi=(T(t,x,u,u_x),X(t,x,u,u_t,u_x,u_{xx}))
\end{equation}
whose form arises from 
splitting the characteristic equation \eqref{gKdV-conslaw}
with respect to $u_{tx}$, $u_{xxx}$, $u_{xxxx}$.  
Moreover, 
since $u_x,u_{xx}$ are the only derivatives of $u$ that can be differentiated
to yield a leading derivative of the gKdV equation \eqref{gKdV-eqn}, 
multipliers with the form \eqref{gKdV-2ndordQ} generate 
generate all low-order conserved currents \eqref{gKdV-TX} 
for the gKdV equation. 

We now apply conditions \eqref{gKdV-Pcond2} and \eqref{gKdV-Pcond1}
to a linear combination of the point symmetry characteristic functions \eqref{gKdv-pointP}. 
This yields the multipliers
\begin{gather}
Q_1 = u_{xx} + \tfrac{k}{p+1}u^{p+1}, 
\quad
Q_2 = -u, 
\label{gKdV-Q}\\
Q_3 = 3t(u_{xx} + ku^3) -xu, 
\quad 
p=2 , 
\label{gKdV-Q-p=2}\\
Q_4 = x-ktu,
\quad 
p=1 , 
\label{gKdV-Q-p=1}
\end{gather}
which can be seen to correspond to the point symmetries 
$\X_1$, $\X_2$, $\X_3$ with $p=2$, and $\X_4$. 
These multipliers are known to comprise 
all solutions of the determining system \eqref{gKdV-adjsymmeq} and \eqref{gKdV-helmhotzeq}
for multipliers of at most second order \cite{AncBlu02a}, 
apart from the obvious constant solution
\begin{equation}\label{gKdV-Q-const}
Q_5=1
\end{equation}
(which can be viewed as arising from a trivial symmetry, $P=0$, 
by the correspondence relation \eqref{gKdV-PQrel}). 
Note we have 
\begin{gather}
R_{Q_5} = 0,
\quad
R_{Q_1} = -D_x{}^2-k u^p, 
\quad
R_{Q_2} = 1,
\label{gKdV-RQ}\\
R_{Q_3} = -3t(D_x{}^2+ku^2) + x,
\quad
p=2 , 
\label{gKdV-RQ-p=2}\\
R_{Q_4} = kt,
\quad
p=1 ,
\label{gKdV-RQ-p=1}
\end{gather}
as given by equation \eqref{deteqoffE}. 

For each multiplier \eqref{gKdV-Q}--\eqref{gKdV-Q-const}, 
a conserved current \eqref{gKdV-TX} can be obtained directly by 
either integration of the characteristic equation \eqref{gKdV-conslaw}
or use of the scaling formula $\Psi_G(3tu_t +xu_x +(2/p)u,Q)$ 
from \propref{PQformula}. 
In particular, the components of the scaling formula are given by 
\begin{equation}
\begin{aligned}
\Psi_G^t & = (\tfrac{2}{p}u+xu_x -3t(u_{xxx}+ku^pu_x))Q, 
\\
\Psi_G^x & = 
(\tfrac{2}{p} u+xu_x -3t(u_{xxx}+ku^pu_x))(ku^p Q + D_x{}^2 Q)
\\&\qquad
-((1+\tfrac{2}{p})u_x+xu_{xx} -3t(u_{xxxx}+k(u^pu_x)_x))D_x Q
\\&\qquad
+(2(1+\tfrac{1}{p})u_{xx}+xu_{xxx} -3t(u_{xxxxx}+k(u^pu_x)_{xx}))Q . 
\end{aligned}
\end{equation}
We obtain, modulo the addition of a trivial current, 
\begin{align}
&\begin{aligned}
T_5 = u, 
\quad
X_5 = u_{xx} + \tfrac{k}{p+1}u^{p+1};
\end{aligned}
\label{gKdV-TX-5}\\
&\begin{aligned}
T_1=  -\tfrac{1}{2}u_x^2 + \tfrac{k}{(p+1)(p+2)}u^{p+2}, 
\quad
X_1 = \tfrac{1}{2}\big(u_{xx} + \tfrac{k}{p+1}u^{p+1}\big)^2 +u_tu_x;
\end{aligned}
\label{gKdV-TX-1}\\
&\begin{aligned}
T_2 = -\tfrac{1}{2} u^2,
\quad
X_2 = -uu_{xx}+\tfrac{1}{2}u_x^2-\tfrac{k}{p+2} u^{p+2};
\end{aligned}
\label{gKdV-TX-2}\\
&\begin{aligned}
& T_3 = -t\big(\tfrac{3}{2} u_x^2 + \tfrac{k}{4} u^4) - \tfrac{1}{2}xu^2,
\\
& X_3 = \tfrac{3}{2} t\big(u_{xx} +\tfrac{k}{3}u^3\big)^2 -x\big( uu_{xx} - \tfrac{1}{2}u_x^2  + \tfrac{k}{4} u^4\big) + (3tu_t+u)u_x,
\quad 
p=2;
\end{aligned}
\label{gKdV-TX-3}\\
&\begin{aligned}
& T_4 = -\tfrac{1}{2} kt u^2 -xu, 
\\
& X_4 =  -kt\big(uu_{xx} -\tfrac{1}{2}u_x^2 - \tfrac{k}{3}u^3\big) +x\big(u_{xx} + \tfrac{k}{2} u^2\big) - u_x,
\quad 
p=1. 
\end{aligned}
\label{gKdV-TX-4}
\end{align}
These five conserved currents respectively represent conservation of 
mass, momentum, energy, Galilean energy, and Galilean momentum 
for the gKdV equation \eqref{gKdV-eqn}. 

We will now study the symmetry properties of the conservation laws 
\eqref{gKdV-TX-5}--\eqref{gKdV-TX-4}. 
Consider the vector space of conserved currents
\begin{equation}\label{gKdV-gencurrent}
\begin{aligned}
& T = a_1 T_1 +a_2 T_2 +a_3 T_3 +a_4 T_4+a_5 T_5,
\\
& a_i=\const,
\quad
a_3 = 0\text{ if }p\neq 2,
\quad
a_4 = 0\text{ if }p\neq 1,
\end{aligned}
\end{equation}
and the algebra of point symmetries
\begin{equation}\label{gKdV-gensymm}
\X = c_1\X_1+c_2 \X_2+c_3 \X_3+c_4 \X_4,
\quad
c_j=\const,
\quad
c_4 = 0\text{ if }p\neq 1 . 
\end{equation}
A conservation law $(D_tT+D_xX)|_\E=0$ is homogeneous under the symmetry $\X$ 
iff condition \eqref{invcond} is satisfied, 
where the characteristic of the symmetry generator is given by 
$P= c_1P_1+c_2P_2+c_3 P_3+c_4P_4$,
and the multiplier for the conservation law is given by 
$Q= a_1Q_1+a_2Q_2+a_3 Q_3 +a_4 Q_4 +a_5 Q_5$. 
By using equations \eqref{gKdV-RP} and \eqref{gKdV-RQ}--\eqref{gKdV-RQ-p=1}, 
we find that the condition \eqref{invcond} splits with respect to $t,x,u,u_x,u_{xx}$
into a system of bilinear equations on $c_j$ and $a_i$:
\begin{subequations}\label{gKdV-aceqs}
\begin{align}
p\neq 1,2:&\quad
a_2(p\lambda +(4-p)c_3) =0
\\
p\neq 1:&\quad
a_5(p\lambda +(2-p)c_3) =0
\\
p\neq 2:&\quad
a_1(p\lambda+(p+4)c_3) =0
\\
p=1:&\quad
a_4\lambda=0,
\quad
k(a_1c_4-a_4c_1)+ a_2(\lambda +3c_3) =0,
\quad
a_2c_4 -a_4c_2+a_5(\lambda +c_3) =0
\\
p=2:&\quad
a_3\lambda=0,
\quad
3a_3c_1-a_1(\lambda+3c_3) =0,
\quad
a_3c_2-a_2(\lambda +c_3) =0
\end{align}
\end{subequations}
The solutions for $c_j$ in terms of $a_i$ 
determine the symmetry-homogeneity properties of 
the conserved current $\Phi=(T,X)$ modulo trivial currents. 
By considering the subspaces generated by $\{a_i\}$,
solving the system \eqref{gKdV-aceqs} in each case, 
and merging the solutions, we get the conditions
\begin{subequations}
\begin{align}
p\neq 1:&\quad
a_5\neq 0, 
\quad
a_1=a_2=a_4=0;
\quad
\lambda = (1-2/p)c_3
\\
p\neq 1:&\quad
a_2\neq 0,
\quad
a_1=a_5=a_3=0;
\quad
\lambda = (1-4/p)c_3
\\
p\neq 1:&\quad
a_1\neq 0,
\quad
a_2=a_5=a_3=0;
\quad
\lambda = -(1+4/p)c_3
\\
p=1:&\quad
a_1{}^2+a_2{}^2+a_5{}^2\neq 0, 
\quad
a_2{}^2= 2k a_1a_5, 
\quad
a_4=0;
\quad
a_2c_4 =4a_5 c_3, 
\quad
\lambda = -5c_3
\\
p=1:&\quad
a_2{}^2+a_5{}^2\neq 0, 
\quad
a_1=a_4=0;
\quad
a_2c_4= 2a_5c_3, 
\quad
\lambda = -3c_3
\\
p=1:&\quad
a_5\neq 0,
\quad
a_1=a_2=a_4=0;
\quad
\lambda = -c_3
\\
p=1:&\quad
a_2\neq 0,
\quad
a_1=a_5=a_4=0;
\quad
c_4 =0,
\quad
\lambda = -3c_3
\\
p=1:&\quad
a_1\neq 0,
\quad
a_2=a_5=a_4=0;
\quad
c_4=0, 
\quad
\lambda = -5c_3
\\
p=1:&\quad
a_1{}^2+a_2{}^2+a_4{}^2+a_5{}^2\neq 0;
\quad
c_3=0,
\quad
a_4c_1=a_1c_4, 
\quad
a_4c_2=a_2c_4, 
\quad
\lambda = 0
\\
p=1:&\quad
a_2{}^2+a_4{}^2+a_5{}^2\neq 0,
\quad
a_1=0;
\quad
ka_4c_1=3a_2c_3, 
\quad
a_4c_2=a_2c_4+a_5c_3, 
\quad
\lambda = 0
\\
p=1:&\quad
a_1{}^2+a_5{}^2+a_4{}^2\neq 0,
\quad
a_2=0;
\quad
c_2=c_3=0,
\quad
a_4c_1=a_1c_4, 
\quad
\lambda = 0
\\
p=2:&\quad
a_1{}^2+a_2{}^2+a_5{}^2+a_3{}^2\neq 0;
\quad
a_3c_1=a_1c_3, 
\quad
a_3c_2=a_2c_3,
\quad
a_1c_2=a_2c_1,
\quad
\lambda = 0
\end{align}
\end{subequations}
Hence, we conclude the following.

(1) 
For $p\neq 1,2$, 
the symmetry properties of the vector space 
$a_1\Phi_1 +a_2\Phi_2+a_5\Phi_5+\Phi_\triv$ are generated by:
(i) invariance under $\X_1,\X_2$;
(ii) homogeneity of the subspace $a_1\Phi_1+\Phi_\triv$ 
under $\X_3$ with $\lambda =-1-4/p$;
(iii) homogeneity of the subspace $a_2\Phi_2+\Phi_\triv$ 
under $\X_3$ with $\lambda =1-4/p$;
(iv) homogeneity of the subspace $a_5\Phi_5+\Phi_\triv$ 
under $\X_3$ with $\lambda =1-2/p$.

(2) 
For $p=1$, 
the symmetry properties of the vector space 
$a_1\Phi_1 +a_2\Phi_2+a_5\Phi_5+a_4\Phi_4+\Phi_\triv$ are generated by:
(i) invariance of the subspace $a_1\Phi_1+a_2\Phi_2+a_4\Phi_4+a_5\Phi_5+\Phi_\triv$
under $a_1\X_1+a_2\X_2 +a_4\X_4$;
(ii) invariance of the subspace $a_2\Phi_2+a_5\Phi_5+a_4\Phi_4+\Phi_\triv$ 
under $a_2\X_2+a_4\X_4$ and $\tfrac{3}{k} a_2\X_1+ a_5\X_2 +a_4\X_3$;
(iii) invariance of the subspace $a_1\Phi_1+a_5\Phi_5+a_4\Phi_4+\Phi_\triv$ 
under $a_1\X_1+a_4\X_4$;
(iv) invariance of the subspace $a_1\Phi_1+\Phi_\triv$ under $\X_2$;
(v) invariance of the subspace $a_5\Phi_5+\Phi_\triv$ under $\X_1$;
(vi) homogeneity of the projective subspace $a_1\Phi_1+a_2\Phi_2+a_5\Phi_5+\Phi_\triv$, $a_2{}^2= 2k a_1a_5$, 
under $a_2\X_3 +4a_5\X_4$ with $\lambda=-5a_2$; 
(vii) homogeneity of the subspace $a_2\Phi_2+a_5\Phi_5+\Phi_\triv$
under $a_2\X_3 +2a_5\X_4$ with $\lambda=-3a_2$; 
(viii) homogeneity of the subspace $a_1\Phi_1+\Phi_\triv$
under $\X_3$ with $\lambda=-5$; 
(ix) homogeneity of the subspace $a_2\Phi_2+\Phi_\triv$
under $\X_3$ with $\lambda=-3$; 
(x) homogeneity of the subspace $a_5\Phi_5+\Phi_\triv$
under $\X_3$ with $\lambda=-1$. 

(3) 
For $p=2$, 
the symmetry properties of the vector space 
$a_1\Phi_1 +a_2\Phi_2+a_5\Phi_5+a_3\Phi_3+\Phi_\triv$ are generated by:
(i) invariance under $a_1\X_1+ a_2\X_2 +a_3\X_3$;
(ii) invariance of the subspace $a_1\Phi_1+\Phi_\triv$ 
under $\X_2$;
(iii) invariance of the subspace $a_2\Phi_2+\Phi_\triv$ under $X_1$;
(iv) invariance of the subspace $a_5\Phi_5+\Phi_\triv$ 
under $\X_1,\X_2,\X_3$;
(v) homogeneity of the subspace $a_1\Phi_1+\Phi_\triv$
under $\X_3$ with $\lambda=-3$; 
(vi) homogeneity of the subspace $a_2\Phi_2+\Phi_\triv$
under $\X_3$ with $\lambda=-1$. 

From these properties, it follows that
the mass, momentum, energy conservation laws \eqref{gKdV-TX-5}--\eqref{gKdV-TX-2} 
admitted for $p\neq 1,2$ 
are homogeneous under the scaling symmetry $\X_3$
and invariant under the translation symmetries $\X_1$ and $\X_2$,
while the Galilean energy conservation law \eqref{gKdV-TX-3}
admitted only for $p=2$ is invariant under $\X_3$
and the Galilean momentum conservation law \eqref{gKdV-TX-4} 
admitted only for $p=1$ is invariant under both $\X_3$ and $\X_4$.
In particular, 
the scaling symmetry $\X_3$ maps 
$\int_\Omega T_5 dx$ into $e^{(1-2/p)\epsilon}\int_\Omega T_5 dx$, 
$\int_\Omega T_1 dx$ into $e^{-(1+4/p)\epsilon}\int_\Omega T_1 dx$, 
and $\int_\Omega T_2 dx$ into $e^{(1-4/p)\epsilon}\int_\Omega T_2 dx$,
if boundary conditions are imposed such that all endpoint terms vanish.  

Additionally, for $p=1$, 
a combined scaling and Galilean boost symmetry $a_2\X_3+4a_5\X_4$ 
maps the conserved quantity $\int_\Omega a_1T_1\pm\sqrt{2k a_1a_5}T_2+a_5T_5 dx$
into $e^{-5a_2\epsilon}\int_\Omega a_1T_1\pm\sqrt{2k a_1a_5}T_2+a_5T_5 dx$,
and a similar symmetry $a_2\X_3+2a_5\X_4$ 
maps the conserved quantity $\int_\Omega a_2T_2+a_5T_5 dx$
into $e^{-3a_2\epsilon}\int_\Omega a_2T_2+a_5T_5 dx$, 
under suitable boundary conditions. 
These symmetry-homogeneous conserved quantities represent 
linear combinations of mass, momentum, and energy. 
Likewise, 
the conserved quantities 
$\int_\Omega a_2T_2 +a_4T_4 +a_1T_1 dx$ and $\int_\Omega a_2T_2+a_4T_4 +a_5T_5 dx$ 
representing linear combinations of momentum, Galilean momentum, and energy or mass 
are invariant under the respective symmetries 
$a_1\X_1+a_2\X_2 +a_4\X_4$ and $a_2\X_1+\tfrac{k}{3}(a_5\X_2 +a_4\X_3)$
representing $t,x$-translation symmetries combined with a Galilean boost or a scaling. 

It is interesting to note that every conserved quantity 
$\int_\Omega a_1T_1 + a_2T_2 +a_3T_3 +a_4T_4 dx$ is invariant 
under the variational symmetry $\hat\X=D_xQ\p_u$ that corresponds to 
its multiplier $Q=a_1Q_1 + a_2Q_2 +a_3Q_3 +a_4Q_4$ 
through the relation \eqref{gKdV-PQrel}. 
This relation also shows that $R_P=-D_xR_Q$ holds, 
from which we see that the symmetry-invariance condition $R_P^*(Q)-R_Q^*(P)=0$
holds identically when $D_xQ=P$. 
As a consequence, 
the symmetry-invariance property stated in \thmref{symmcond} 
for conservation laws of Euler-Lagrange PDEs 
extends to the case of PDEs that have the form of Euler-Lagrange equations
when a potential is introduced. 

It is also interesting that, 
for the Lagrangian formulation of the gKdV equation, 
homogeneity under the non-variational scaling symmetry \eqref{gKdV-pt3} 
when $p\neq 1,2$ determines three conservation laws, 
in contrast to Noether’s theorem which cannot yield a conservation law 
from this symmetry when $p\neq 1,2$. 

Finally, 
we remark that the correspondence relation $D_xQ=P$
between multipliers and variational symmetries 
can also be derived in a general way from the well-known 
Hamiltonian formulation of the gKdV equation \eqref{gKdV-eqn},
$u_t= D_x(\delta H/\delta u)$
where $H=-\int_\Rnum T_1dx$ is the Hamiltonian 
and $D_x$ is a Hamiltonian operator. 
In particular, 
a general result in the theory of Hamiltonian PDEs \cite{Olv} 
shows that every conserved density $T$ admitted by a PDE of the form 
$u_t= D_x(\delta H/\delta u)$ 
yields a corresponding Hamiltonian symmetry whose characteristic function is 
simply $P=D_x Q$ with $Q=E_u(T)$. 

We also mention that the bi-Hamiltonian structures \cite{Olv} of 
the KdV equation and the modified KdV equation 
can be used to show that all of the higher-order conservation laws 
for these integrable equations are invariant under 
all of the higher-order symmetries generated from $\hat\X=u_x\p_u$
by the recursion operator for each equation.

\subsection{Nonlinear viscous fluid equation}
\label{gB-ex}

Our third example is the PDE
\begin{equation}\label{gB-eqn}
u_t +uu_x= k(u^pu_x)_x,
\quad
p\neq 0, k\neq 0
\end{equation}
which reduces to Burgers' equation when $p=1$. 
This nonlinear PDE \eqref{gB-eqn} models a non-Newtonian viscous fluid 
where, for $p\neq 1$,  
the viscosity coefficients depend nonlinearly on the fluid velocity \cite{Bat}.
Note that either $u_{xx}$ or $u_t$ is a leading derivative, so the PDE is normal. 
We will refer to it as a generalized non-Newtonian Burgers' (gnNB) equation. 

The determining equation \eqref{symmdeteq} 
for infinitesimal symmetries $\X=P\p_u$ 
of the gnNB equation \eqref{gB-eqn} is given by 
\begin{equation}\label{gB-symmeq}
(D_tP +D_x(uP) - kD_x{}^2(u^pP))|_\E=0 . 
\end{equation}
Point symmetries and contact symmetries are given by characteristic functions 
$P(t,x,u,u_t,u_x)$. 
The computation of these symmetries is simplest when 
the leading derivative $u_{xx}=\tfrac{1}{k}u^{-p}(u_t+uu_x) -pu^{-1}u_x^2$ is used
(as otherwise $u_t=-uu_x +k(u^pu_x)_x$ must be eliminated from $P$). 
This yields 
\begin{equation}
P_1 = -u_t, 
\quad
P_2 = -u_x, 
\quad
P_3 = -(p-2)tu_t-(p-1)xu_x + u, 
\end{equation}
from which we obtain
\begin{equation}\label{gB-RP}
R_{P_1} = -D_t,
\quad
R_{P_2} = -D_x,
\quad
R_{P_3} = (2-p)tD_t +(1-p)xD_x + (3-p)
\end{equation}
as given by equation \eqref{deteqoffE}. 
These symmetries are point symmetries,
which consist of separate translations in $t$ and $x$, 
and a scaling in $t,x,u$:
\begin{equation}\label{gB-pointsymms}
\X_1 = \p_t, 
\quad
\X_2 = \p_x, 
\quad
\X_3 = (p-2)t\p_t +(p-1)x\p_x +  u\p_u . 
\end{equation}
Their respective group actions consist of 
\begin{align}
& t\rightarrow t+\epsilon,
\label{gnNB-pt1}\\
& x\rightarrow x+\epsilon,
\label{gnNB-pt2}\\
& t\rightarrow e^{(p-2)\epsilon}t,
\quad
x\rightarrow e^{(p-1)\epsilon}x,
\quad
u\rightarrow e^{\epsilon}u 
\label{gnNB-pt3}
\end{align}
in terms of a parameter $\epsilon$. 

The characteristic equation for conservation laws of the gnNB equation \eqref{gB-eqn} is given by 
\begin{equation}\label{gB-conslaw}
D_t T + D_x X = QG,
\quad
G= u_t +uu_x-k(u^pu_x)_x
\end{equation}
where it is simplest and physically natural to eliminate all $t$-derivatives of $u$ from $T$
by using $u_t=-uu_x +k(u^pu_x)_x$ as the leading derivative. 
Then the resulting multipliers $Q$ are functions only of $t,x,u$ and $x$-derivatives of $u$.
Low-order conservation laws are defined by multipliers with the first-order form
$Q(t,x,u,u_x)$
where $u_t$ is excluded because it cannot be differentiated to obtain a leading derivative $u_t$ or $u_{xx}$ of the gnNB equation. 
The corresponding conserved currents will have the form 
\begin{equation}\label{gB-lowordTX}
\Phi=(T(t,x,u),X(t,x,u)u_x)
\end{equation}
which can be readily derived by splitting equation \eqref{gB-conslaw} 
with respect to $u_t$ and $u_{xx}$.
This splitting also yields the relation 
$Q = E_{u}(T) = -\tfrac{1}{k}u^{-p}E_{u_x}(X)$
from which $Q_{u_x}=0$ is obtained through expression \eqref{gB-lowordTX} for $T$ and $X$. 
Hence, multipliers for low-order conservation laws will have the form 
\begin{equation}\label{gB-lowordQ}
Q(t,x,u) . 
\end{equation}

The determining system \eqref{adjsymmdeteq} and \eqref{helmholtzeq} 
for all low-order conservation laws \eqref{gB-lowordTX} consists of 
only the adjoint-symmetry equation 
\begin{equation}\label{gB-adjsymmeq}
-(D_tQ +uD_xQ +ku^pD_x{}^2 Q)|_\E=0
\end{equation}
since the Helmholtz equations \eqref{helmholtzeq} can be shown to hold identically 
due to $Q$ not containing any derivatives of $u$. 
As a consequence, 
low-order adjoint-symmetries of the gnNB equation \eqref{gB-eqn} 
are the same as low-order multipliers. 
There is no correspondence between multipliers and symmetries for this equation.
We also remark that the lack of dependence on derivatives of $u$ in the multipliers \eqref{gB-lowordQ}
is well-known to hold more generally for any quasi-linear second order evolution equation.

A straightforward computation of adjoint-symmetries \eqref{gB-lowordQ}
yields the multipliers 
\begin{gather}
Q_1 = 1, 
\label{gB-Q}\\
Q_2 = e^{-x/k}, 
\quad 
p=1, 
\label{gB-Q-p=1}
\end{gather}
with 
\begin{equation}\label{gB-RQ}
R_{Q_1} = 0,
\quad
R_{Q_2} = 0
\end{equation}
given by equation \eqref{adjsymmdeteqoffE}. 
For each multiplier \eqref{gB-Q}--\eqref{gB-Q-p=1}, 
a conserved current \eqref{gB-lowordTX} can be obtained directly by 
either integration of the characteristic equation \eqref{gB-conslaw}
or use of the scaling formula $\Psi_G(u-(p-2)tu_t-(p-1)xu_x,Q)$
from \propref{PQformula}. 
In particular, the components of the scaling formula are given by 
\begin{equation}
\begin{aligned}
\Psi_G^t & = (u-(p-2)tu_t-(p-1)xu_x)Q, 
\\
\Psi_G^x & = u\Psi_G^t  
-(u^{p+1}-(p-2)tu^pu_t-(p-1)xu^pu_x) D_x Q
\\&\qquad
+(2u^pu_x-(p-2)t(u^pu_t)_x-(p-1)x(u^pu_x)_x) Q . 
\end{aligned}
\end{equation}
We obtain, modulo the addition of a trivial current, 
\begin{align}
&\begin{aligned}
T_1 = u, 
\quad
X_1 = \tfrac{1}{2}u^2 -ku^pu_{xx} ;
\end{aligned}
\label{gB-TX-1}\\
&\begin{aligned}
T_2 =  e^{-x/k} u,
\quad
X_2 = -ke^{-x/k} uu_x,
\quad
p=1 .
\end{aligned}
\label{gB-TX-2}
\end{align}
These two conserved currents respectively represent conservation of 
momentum and exponentially-weighted momentum 
for the gnNB equation \eqref{gB-eqn}. 

We will now study the symmetry properties of the conservation laws 
\eqref{gB-TX-1}--\eqref{gB-TX-2}. 
Consider the vector space of conserved currents
\begin{equation}\label{gB-gencurrent}
T = a_1 T_1 +a_2 T_2, 
\quad
a_i=\const,
\quad
a_2 = 0\text{ if }p\neq 1, 
\end{equation}
and the algebra of point symmetries
\begin{equation}\label{gB-gensymm}
\X = c_1\X_1+c_2 \X_2+c_3 \X_3, 
\quad
c_j=\const . 
\end{equation}
A conservation law $(D_tT+D_xX)|_\E=0$ is homogeneous under the symmetry $\X$ 
iff condition \eqref{invcond} is satisfied, 
where the characteristic of the symmetry generator is given by 
$P= c_1P_1+c_2P_2+c_3 P_3$,
and the multiplier for the conservation law is given by 
$Q= a_1Q_1+a_2Q_2$. 
By using equations \eqref{gB-RP} and \eqref{gB-RQ}, 
we find that the condition \eqref{invcond} splits with respect to $e^{-x/k}$
into a system of bilinear equations on $c_j$ and $a_i$:
\begin{subequations}\label{gB-aceqs}
\begin{align}
p\neq 1:&\quad
pc_3-\lambda=0
\\
p=1:&\quad
a_1(c_3-\lambda) =0,
\quad
a_2(c_2+k(\lambda -c_3)) =0
\end{align}
\end{subequations}
The solutions for $c_j$ in terms of $a_i$ 
determine the symmetry-homogeneity properties of 
the conserved current $\Phi=(T,X)$ modulo trivial currents. 
Solving the system \eqref{gB-aceqs}, we get the conditions
\begin{subequations}
\begin{align}
p\neq 1:&\quad
\lambda = pc_3
\\
p=1:&\quad
a_1{}^2+a_2{}^2 \neq 0,
\quad
c_2 =0, 
\quad
\lambda = c_3
\\
p=1:&\quad
a_1{}^2 \neq 0,
\quad
a_2=0;
\quad
\lambda = c_3
\\
p=1:&\quad
a_2{}^2 \neq 0,
\quad
a_1=0;
\quad
\lambda = c_3-\tfrac{1}{k}c_2
\end{align}
\end{subequations}
Hence, we conclude the following.

(1) 
For $p\neq 1$, 
the symmetry properties of the vector space 
$a_1\Phi_1 +\Phi_\triv$ consist of: 
(i) invariance under $\X_1,\X_2$;
(ii) homogeneity under $\X_3$ with $\lambda =p$. 

(2) 
For $p=1$, 
the symmetry properties of the vector space 
$a_1\Phi_1 +a_2\Phi_2+\Phi_\triv$ for arbitrary $a_i$ consist of:
(i) invariance under $\X_1$;
(ii) homogeneity under $\X_3$ with $\lambda=1$.

(3) For $p=1$, the only additional symmetry properties of the vector space 
$a_1\Phi_1 +a_2\Phi_2+\Phi_\triv$ consist of:
(i) invariance of the subspace $a_1\Phi_1+\Phi_\triv$ 
under $\X_2$;
(ii) homogeneity of the subspace $a_2\Phi_2+\Phi_\triv$
under $\X_2$ with $\lambda=-\tfrac{1}{k}$. 

From these properties, it follows that
the momentum conservation law \eqref{gB-TX-1} admitted for any $p$ 
is homogeneous under the scaling symmetry $\X_3$
and invariant under the translation symmetries $\X_1$ and $\X_2$,
while the exponentially-weighted momentum conservation law \eqref{gB-TX-2} 
admitted only for $p=1$ is homogeneous under both the scaling symmetry $\X_3$
and the $x$-translation symmetry $\X_2$, 
and is invariant under the $t$-translation symmetry $\X_1$
as well as the combined symmetry $\X_3+k\X_2$. 
In particular, 
the scaling symmetry $\X_3$ maps 
$\int_\Omega T_1 dx$ into $e^{p\epsilon}\int_\Omega T_1 dx$, 
and $\int_\Omega T_2 dx$ into $e^{\epsilon}\int_\Omega T_2 dx$,
if boundary conditions are imposed such that all endpoint terms vanish.

\subsection{Peakon equation}
\label{bfam-ex}

Our fourth example is the $b$-family peakon equation 
\begin{equation}\label{bfam-sys}
m_t +u m_x +bu_x m,
\quad
m=u-u_{xx},
\quad
b\neq -1
\end{equation}
which arises \cite{DulGotHol} from the theory of shallow water waves
and includes the Camassa-Holm equation \cite{CamHol} when $b=2$
and the Degasperis-Procesi equation \cite{DegPro} when $b=3$. 
As shown in \Ref{DegHonHol}, 
the $b$-family equation possesses multi-peakon solutions 
and has a Hamiltonian structure 
\begin{equation}\label{bfam-Ham}
m_t = \Dop(\delta H/\delta m), 
\quad
H= \begin{cases}
(b-1)^{-1} \int_\Rnum m\;dx, & b\neq 1
\\
\int_\Rnum m\ln(m)\;dx, & b=1
\end{cases}
\end{equation}
given in terms of the Hamiltonian operator 
\begin{equation}\label{bfam-Hamop}
\Dop = (m^{1-1/b}D_x m^{1/b})(D_x\Delta)^{-1}(m^{1/b}D_x m^{1-1/b}) . 
\end{equation}

For studying conservation law and symmetries, 
it is more useful to work with the equivalent PDE 
\begin{equation}\label{bfam-eqn}
u_t -u_{txx} + (b+1)uu_x = bu_xu_{xx} + uu_{xxx},
\quad
b\neq -1 .
\end{equation}
Note that $u_{xxx}$ or $u_{txx}$ is a leading derivative in this $b$-family equation. 
Since $u_{xxx}$ is a pure derivative, the $b$-family equation is a normal PDE. 

The determining equation \eqref{symmdeteq} 
for infinitesimal symmetries $\X=P\p_u$ 
of the $b$-family equation \eqref{bfam-eqn} is given by 
\begin{equation}\label{bfam-symmeq}
( D_t\Delta P +u D_x\Delta P +b u_x \Delta P
+b(u-u_{xx})D_x P + (u_x-u_{xxx})P )|_\E=0 
\end{equation}
with $\Delta=1-D_x{}^2$. 
A direct calculation shows that the $b$-family equation 
has no contact symmetries 
and that all of its point symmetries are generated by 
separate translations in $t$ and $x$, a scaling in $t,u$, 
and, when $b=0$, a Galilean boost.  
For these symmetries
\begin{gather}
\X_1 = \p_t, 
\quad
\X_2 = \p_x, 
\quad
\X_3 = u\p_u -t\p_t , 
\label{bfam-pointsymms}
\\
\X_4 = \p_u +t\p_x,
\quad
b=0 , 
\label{bfam-pointsymms-p=2}
\end{gather}
the characteristic functions are given by 
\begin{gather}
P_1 = -u_t, 
\quad
P_2 = -u_x, 
\quad
P_3 = u+t u_t,
\\
P_4 = 1-tu_x,
\quad
b=0 ,
\end{gather}
from which we obtain
\begin{equation}\label{bfam-RP}
R_{P_1} = -D_t,
\quad
R_{P_2} = -D_x,
\quad
R_{P_3} = tD_t +2,
\quad
R_{P_4} = -tD_x,
\end{equation}
as given by equation \eqref{deteqoffE}. 
Their respective group actions are given by 
\begin{align}
& t\rightarrow t+\epsilon,
\label{bfam-pt1}\\
& x\rightarrow x+\epsilon,
\label{bfam-pt2}\\
& t\rightarrow e^{-p\epsilon}t,
\quad
u\rightarrow e^{\epsilon}u,  
\label{bfam-pt3}\\
& x\rightarrow x +\epsilon t, 
\quad
u\rightarrow u + \epsilon , 
\label{bfam-pt4}
\end{align}
in terms of a parameter $\epsilon$. 
(A symmetry classification presented in \Ref{SinGupKum} is missing the case \eqref{bfam-pointsymms-p=2}.)

To formulate the characteristic equation for conservation laws of the $b$-family equation \eqref{bfam-eqn},
it is simplest to use the solved form 
$u_{xxx}=u_x + u^{-1}(bu_x(u-u_{xx})- u_t+u_{txx})$
in terms of the leading derivative $u_{xxx}$. 
This yields 
\begin{equation}\label{bfam-conslaw}
D_t T + D_x X = QG,
\quad
Q = -u^{-1}E_{u_{xx}}(X),
\quad
G= u_t -u_{txx} + (b+1)uu_x -bu_xu_{xx} - uu_{xxx}
\end{equation}
where $X$ is a function of $t,x,u,u_x,u_{xx}$, and $t$-derivatives of $u,u_x,u_{xx}$,
with all $x$-derivatives of $u_{xx}$ being eliminated from $X$. 
Consequently, 
multipliers $Q$ are also functions only of $t,x,u,u_x,u_{xx}$, 
and $t$-derivatives of $u,u_x,u_{xx}$. 
Low-order conservation laws are defined by multipliers with the second-order form 
\begin{equation}\label{bfam-lowordQ}
Q(t,x,u,u_t,u_x,u_{tx},u_{xx})  
\end{equation}
where $u_{tt}$ is excluded because it cannot be differentiated to obtain a leading derivative $u_{xxx}$ or $u_{txx}$ of the $b$-family equation. 
The corresponding conserved currents will have the form 
\begin{equation}\label{bfam-lowordTX}
\Phi=(T(t,x,u,u_x,u_{xx}),X(t,x,u,u_t,u_x,u_{tx},u_{xx}))
\end{equation}
which can be readily derived by splitting equation \eqref{bfam-conslaw} 
with respect to $u_{tt}$ and its differential consequences.

The determining system \eqref{adjsymmdeteq} and \eqref{helmholtzeq} 
for all low-order conservation laws \eqref{bfam-lowordTX} consists of 
only the adjoint-symmetry equation 
\begin{equation}\label{bfam-adjsymmeq}
( -D_t\Delta Q -D_x\Delta(u Q) +b\Delta(u_x Q) -bD_x((u-u_{xx})Q) 
+(u_x-u_{xxx})Q )|_\E=0 
\end{equation}
since the Helmholtz equations \eqref{helmholtzeq} can be shown to hold identically 
due to $Q$ not containing any $t$-derivatives of $u_{xx}$. 
Hence, 
low-order adjoint-symmetries of the $b$-family equation \eqref{bfam-eqn} 
are the same as low-order multipliers. 
There is no correspondence between multipliers 
$Q(t,x,u,u_x,u_{xx},u_t,u_{tx},u_{txx},\ldots)$ 
and symmetry characteristics $P(t,x,u,u_x,u_{xx},u_t,u_{tx},u_{txx},\ldots)$ 
for this equation.
(However, if the class of multipliers and symmetries is enlarged 
by including certain nonlocal variables that arise from the Hamiltonian structure, then a correspondence within this larger class does exist.)

A straightforward computation of adjoint-symmetries \eqref{bfam-lowordQ}
yields the multipliers 
\begin{gather}
Q_1 = (u-u_{xx})^{-1+1/b},
\quad
b\neq 0
\label{bfam-Q-b<>0}\\
Q_2 = 1
\label{bfam-Q-b<>1}\\
\begin{gathered}
Q_3 = u, 
\quad 
b=2
\\
Q_4= u(u-u_{xx})+\tfrac{1}{2}(u^2-u_x^2)-u_{tx},
\quad 
b=2
\end{gathered}
\label{bfam-Q-b=2}\\
\begin{gathered}
Q_5 = u(u-u_{xx})+\tfrac{1}{2}u^2 -u_x^2-u_{tx},
\quad 
b=3
\\
Q_{6^\pm}= e^{\pm 2x}, 
\quad 
b=3
\end{gathered}
\label{bfam-Q-b=3}\\
Q_7= 1+\ln(u-u_{xx}),
\quad
b=1
\label{bfam-Q-b=1}
\end{gather}
with 
\begin{gather}
R_{Q_1} = (1-1/b)((u-u_{xx})^{-2+1/b}\Delta -2((u-u_{xx})^{-2+1/b})_xD_x),
\quad
b\neq 0
\label{bfam-RQ-b<>0}\\
R_{Q_2} = 0
\label{bfam-RQ-b<>1}\\
\begin{gathered}
R_{Q_3} = -1, 
\quad 
b=2
\\
R_{Q_4}= D_tD_x +D_x(uD_x) -3u+u_{xx}, 
\quad 
b=2
\end{gathered}
\label{bfam-RQ-b=2}\\
\begin{gathered}
R_{Q_5} = D_tD_x +uD_x{}^2 -3u, 
\quad 
b=3
\\
R_{Q_{6^\pm}} = 0,
\quad 
b=3
\end{gathered}
\label{bfam-RQ-b=3}\\
R_{Q_7} = -(u-u_{xx})^{-1}\Delta +2((u-u_{xx})^{-1})_xD_x + ((u-u_{xx})^{-1})_{xx},
\quad 
b=1
\label{bfam-RQ-b=1}
\end{gather}
given by equation \eqref{adjsymmdeteqoffE}. 
For each multiplier \eqref{bfam-Q-b<>0}--\eqref{bfam-Q-b=1}, 
a conserved current \eqref{bfam-lowordTX} can be obtained directly by 
either integration of the characteristic equation \eqref{bfam-conslaw}
or use of the scaling formula $\Psi_G(u+ptu_t,Q)$
from \propref{PQformula}. 
In particular, the components of the scaling formula are given by 
\begin{equation}
\begin{aligned}
\Psi_G^t & = (u-u_{xx}+t(u_t-u_{txx}))Q,
\\
\Psi_G^x & = 
(u+tu_t)(D_tD_xQ -D_x^2(uQ) +bD_x(u_xQ)+b(u-u_{xx})Q)
\\&\qquad
+(u_x+tu_{tx})(-D_tQ+D_x(uQ) -bu_xQ) +u(u-u_{xx}+t(u_t-u_{txx}))Q . 
\end{aligned}
\end{equation}
We obtain, modulo the addition of a trivial current and an overall scaling, 
\begin{align}
&\begin{aligned}
T_1 = (u-u_{xx})^{1/b}, 
\quad
X_1 = -bu(u-u_{xx})^{1/b},
\quad
b\neq 0;
\end{aligned}
\label{bfam-TX-1}\\
&\begin{aligned}
T_2 =  u,
\quad
X_2 = \tfrac{1}{2}(b-1)(u^2-u_x^2) +u(u-u_{xx})-u_{tx} ;
\end{aligned}
\label{bfam-TX-2}\\
&\begin{aligned}
T_3 = \tfrac{1}{2}(u^2+u_x^2),
\quad
X_3 = u(u-u_{xx}-u_{tx}),
\quad
b=2 ;
\end{aligned}
\label{bfam-TX-3}\\
&\begin{aligned}
T_4 =  \tfrac{1}{2}u(u^2+u_x^2),
\quad
X_4 = \tfrac{1}{2}(u_{tx}+u(u_{xx}+\tfrac{3}{2}u)+\tfrac{1}{2}u_x^2)^2 -u_t(uu_x+\tfrac{1}{2}u_t), 
\quad
b=2 ;
\end{aligned}
\label{bfam-TX-4}\\
&\begin{aligned}
T_5 =  \tfrac{1}{2}u^3, 
\quad
X_5 = \tfrac{1}{2}(u_{tx}+u(u_{xx}+\tfrac{3}{2}u)+u_x^2)^2 -\tfrac{1}{2}(u_t+uu_x)^2+\tfrac{3}{8}u^4, 
\quad
b=3 ;
\end{aligned}
\label{bfam-TX-5}\\
&\begin{aligned}
T_{6^\pm} =  e^{\pm 2x}u, 
\quad
X_{6^\pm} = \tfrac{1}{3}e^{\pm 2x}( (u\mp u_x)^2 -u(u-u_{xx})\mp 2u_x +u_{tx}), 
\quad
b=3 ;
\end{aligned}
\label{bfam-TX-6}\\
&\begin{aligned}
T_7 = (u-u_{xx})\ln(u-u_{xx}),
\quad
X_7 = u(u-u_{xx})\ln(u-u_{xx})+\tfrac{1}{2}(u^2-u_x^2),
\quad
b=1 .
\end{aligned}
\label{bfam-TX-7}
\end{align}
The first conserved current is a Hamiltonian Casimir,
and the second conserved current represents conservation of mass
for the $b$-family equation \eqref{bfam-eqn}. 
Note these two currents coincide (modulo a trivial current) when $b=1$. 
The third and fourth conserved currents respectively represent conservation of 
energy and momentum for the Camassa-Holm equation,
while the fifth and sixth conserved currents respectively represent conservation of 
momentum and exponentially-weighted mass for the Degasperis-Procesi equation. 
The seventh conserved current represents conservation of energy. 

We will now study the symmetry properties of these conservation laws 
\eqref{bfam-TX-1}--\eqref{bfam-TX-7}. 
Consider the vector space of conserved currents
\begin{equation}\label{bfam-gencurrent}
\begin{aligned}
& T = a_1 T_1 +a_2 T_2 +a_3 T_3 +a_4 T_4 +a_5 T_5 +a_{6^+} T_{6^+} +a_{6^-} T_{6^-}+a_7 T_7,
\\
& a_i=\const,
\quad
a_1=0\text{ if }b=0,1,
\quad
a_3 =a_4= 0\text{ if }b\neq 2, 
\\&\qquad
a_5 =a_6= 0\text{ if }b\neq 3, 
\quad
a_7 = 0\text{ if }b\neq 1,
\end{aligned}
\end{equation}
and the algebra of point symmetries
\begin{equation}\label{bfam-gensymm}
\X = c_1\X_1+c_2 \X_2+c_3 \X_3 +c_4 \X_4, 
\quad
c_j=\const,
\quad
c_4 =0 \text{ if }b\neq 0 .
\end{equation}
A conservation law $(D_tT+D_xX)|_\E=0$ is homogeneous under the symmetry $\X$ 
iff condition \eqref{invcond} is satisfied, 
where the characteristic of the symmetry generator is given by 
$P= c_1P_1+c_2P_2+c_3 P_3 +c_4 P_4$, 
and the multiplier for the conservation law is given by 
$Q= a_1Q_1+a_2Q_2 +a_3Q_3 +a_4Q_4 +a_5Q_5 +a_{6^+}Q_{6^+}+a_{6^-}Q_{6^-} +a_7Q_7$. 
By using equations \eqref{bfam-RP} and \eqref{bfam-RQ-b<>0}--\eqref{bfam-RQ-b=1}, 
we find that the condition \eqref{invcond} splits with respect to 
$u,u_t,u_x,u_{tx},(u-u_{xx})^{-1+1/b}$ 
into a system of bilinear equations on $c_j$ and $a_i$:
\begin{subequations}\label{bfam-aceqs}
\begin{align}
b\neq 0,1,2,3:&\quad
a_2(c_3-\lambda)=0,
\quad
a_1c_3=0,
\quad
a_1\lambda=0
\\
b=0:&\quad
c_3-\lambda =0
\\
b=1:&\quad
a_7(c_3-\lambda) =0,
\quad
(a_7-a_2)c_3+a_2\lambda=0
\\
b=2:&\quad
a_1(c_3-2\lambda)=0,
\quad
a_2(c_3-\lambda)=0,
\quad
a_3(2c_3-\lambda)=0,
\quad
a_4(3c_3-\lambda)=0
\\
b=3:&\quad
a_1(c_3-3\lambda)=0,
\quad
a_2(c_3-\lambda)=0,
\quad
a_5(3c_3-\lambda)=0,
\quad
a_{6^\pm}(c_3\pm 2c_2-\lambda)=0
\end{align}
\end{subequations}
The solutions for $c_j$ in terms of $a_i$ 
determine the symmetry-homogeneity properties of 
the conserved current $\Phi=(T,X)$ modulo trivial currents. 
Solving the system \eqref{bfam-aceqs}, we get the conditions
\begin{subequations}
\begin{align}
b\neq 0,1,2,3:&\quad
a_2\neq 0,
\quad
a_1=0;
\quad
\lambda = c_3
\\
&\quad
a_1\neq 0,
\quad
a_2=0;
\quad
c_3=0,
\quad
\lambda = 0
\\
b=0:&\quad
a_2\neq 0;
\quad
\lambda = c_3
\\
b=1:&\quad
a_2\neq 0,
\quad
a_7=0;
\quad
\lambda = c_3
\\
b=1:&\quad
a_7\neq 0,
\quad
a_2=0;
\quad
c_3=0,
\quad
\lambda = 0
\\
b=2:&\quad
a_4\neq 0,
\quad
a_2=a_1=a_3=0;
\quad
\lambda = 3c_3
\\
b=2:&\quad
a_3\neq 0,
\quad
a_2=a_1=a_4=0;
\quad
\lambda = 2c_3
\\
b=2:&\quad
a_2{}^2 + a_1{}^2+a_3{}^2+a_4{}^2\neq 0;
\quad
c_3=0,
\quad
\lambda = 0
\\
b=3:&\quad
a_5\neq 0,
\quad
a_2=a_1=a_{6^+}=a_{6^-}=0;
\quad
\lambda = 3c_3
\\
b=3:&\quad
a_{6^\pm}\neq 0,
\quad
a_2=a_1=a_5=a_{6^\mp}=0;
\quad
\lambda = c_3\pm 2c_2
\\
b=3:&\quad
a_2^2+a_{6^\pm}^2\neq 0,
\quad
a_1=a_5=a_{6^\mp}=0;
\quad
c_2=0,
\quad
\lambda = c_3
\\
b=3:&\quad
a_1^2+a_{6^\pm}^2\neq 0,
\quad
a_2=a_5=a_{6^\mp}=0;
\quad
c_3=\mp 3c_2, 
\quad
\lambda = \mp c_2
\\
b=3:&\quad
a_5^2+a_{6^\pm}^2\neq 0,
\quad
a_2=a_1=a_{6^\mp}=0;
\quad
c_3=\mp c_2, 
\quad
\lambda = 3c_3
\\
b=3:&\quad
a_2^2+a_1^2+a_5^2+a_{6^\pm}^2\neq 0;
\quad
c_2=c_3=0,
\quad
\lambda = 0
\end{align}
\end{subequations}
Hence, we conclude the following.

(1) 
For $b\neq 0$, 
the symmetry properties of the vector space $a_1\Phi_1 +a_2\Phi_2+\Phi_\triv$ 
consist of:
(i) invariance under $\X_1,\X_2$;
(ii) homogeneity of the subspace $a_2\Phi_2 +\Phi_\triv$ 
under $\X_3$ with $\lambda=1$. 

(2) 
For $b=0$, 
the symmetry properties of the vector space 
$a_2\Phi_2 +\Phi_\triv$ consist of:
(i) invariance under $\X_1,\X_2,\X_4$;
(ii) homogeneity under $\X_3$ with $\lambda=1$. 

(3) 
For $b=1$, 
the symmetry properties of the vector space 
$a_2\Phi_2 +a_7\Phi_7 +\Phi_\triv$ consist of 
invariance under $\X_1,\X_2$. 

(4)
For $b=2$, 
the symmetry properties of the vector space 
$a_1\Phi_1 +a_2\Phi_2+a_3\Phi_3 +a_4\Phi_4+\Phi_\triv$ consist of:
(i) invariance under $\X_1,\X_2$;
(ii) homogeneity of the subspace $a_3\Phi_3+\Phi_\triv$
under $\X_3$ with $\lambda=2$;
(iii) homogeneity of the subspace $a_4\Phi_4+\Phi_\triv$
under $\X_3$ with $\lambda=3$. 

(5) 
For $b=3$, 
the symmetry properties of the vector space 
$a_1\Phi_1 +a_2\Phi_2+a_5\Phi_5 +a_{6^+}\Phi_{6^+}+a_{6^-}\Phi_{6^-}+\Phi_\triv$ 
consist of:
(i) invariance under $\X_1$;
(ii) invariance of the subspace $a_5\Phi_5+\Phi_\triv$
under $\X_2$;
(iii) invariance of the subspaces $a_{6^\pm}\Phi_{6^\pm}+\Phi_\triv$ 
under $\X_3\mp 2\X_2$;
(iv) homogeneity of the subspace $a_5\Phi_5+\Phi_\triv$
under $\X_3$ with $\lambda=3$;
(v) homogeneity of the subspace $a_{6^+}\Phi_{6^+}+a_{6^-}\Phi_{6^-}+\Phi_\triv$ 
under $\X_3$ with $\lambda=1$;
(vi) homogeneity of the subspaces $a_{6^\pm}\Phi_{6^\pm-}+\Phi_\triv$ 
under $\X_2$ with $\lambda=\pm 2$;
(vii) homogeneity of the subspaces $a_1\Phi_1+a_{6^\pm}\Phi_{6^\pm}+\Phi_\triv$ 
under $3\X_3\mp\X_2$ with $\lambda=1$;
(viii) homogeneity of the subspaces $a_5\Phi_5+a_{6^\pm}\Phi_{6^\pm}+\Phi_\triv$ 
under $\X_3\pm\X_2$ with $\lambda=3$. 

From these properties, it follows that
each of the conservation laws \eqref{bfam-TX-1}--\eqref{bfam-TX-6} 
is homogeneous under the scaling symmetry $\X_3$. 
Additionally, 
the mass conservation law \eqref{bfam-TX-1} admitted for all $b$ 
and the Hamiltonian Casimir conservation law \eqref{bfam-TX-2} 
admitted for any $b\neq0$ 
are invariant under the translation symmetries $\X_1$ and $\X_2$. 
Similarly, 
the energy and momentum conservation laws \eqref{bfam-TX-3}--\eqref{bfam-TX-4}
admitted for $b=2$ (Camassa-Holm equation) 
and the momentum conservation law \eqref{bfam-TX-5}
admitted for $b=3$ (Degasperis-Procesi equation) 
are invariant under the space-time translation symmetries $\X_1$ and $\X_2$,
while the exponentially-weighted mass conservation law \eqref{bfam-TX-6} 
admitted for $b=3$ 
is only invariant under the time translation symmetry $\X_1$
but is homogeneous under the space translation symmetry $\X_2$. 

It is interesting to compare 
the formulation of conservation laws and symmetries for the $b$-family \eqref{bfam-eqn}
and the corresponding formulation for the equivalent system \eqref{bfam-sys}. 
Note the $b$-family system \eqref{bfam-sys} is a normal PDE system 
whose set of leading derivatives is $\{m_x,u_{xx}\}$. 
The corresponding solved form for the system is given by 
\begin{align}
& m_x= -u^{-1}(m_t +bu_x m), 
\label{bfam-sys-m}\\
& u_{xx}=u-m . 
\label{bfam-sys-u}
\end{align}

The determining equation \eqref{symmdeteq} 
for infinitesimal symmetries $\X=P^m\p_m +P^u\p_u$ 
of the $b$-family system \eqref{bfam-sys-m}--\eqref{bfam-sys-u} is given by 
\begin{gather}
( D_x P^m + u^{-1}(D_tP^m +b u_x P^m +b m D_x P^u) -u^{-2}(m_t+b u_xm)P^u )|_\E=0, 
\label{bfam-sys-m-symmeq}\\
(P^m-\Delta P^u)|_\E=0 , 
\label{bfam-sys-u-symmeq}
\end{gather}
where $P^m$ and $P^u$ are functions of $t,x,u,u_x,m$ and $t$-derivatives of $u,u_x,m$. 
If $P^m$ is eliminated in terms of $P^u$ through equation \eqref{bfam-sys-u-symmeq}, 
then equation \eqref{bfam-sys-m-symmeq} simplifies to 
the symmetry determining equation \eqref{bfam-symmeq} with $P=P^u|_{m=u-u_{xx}}$. 
In particular, 
the point symmetries \eqref{bfam-pointsymms}--\eqref{bfam-pointsymms-p=2} 
of the $b$-family equation 
correspond to the characteristic functions
\begin{gather}
P^m_1 = -m_t,
\quad
P^u_1 = -u_t 
\\
P^m_2 = u^{-1}(m_t+bmu_x)= -m_x,
\quad 
P^u_2 = -u_x 
\\
P^m_3 = m+t m_t,
\quad
P^u_3 = u+t u_t
\\
P^m_4 = 1 +tu^{-1}m_t=1-tm_x,
\quad
P^u_4 = 1-tu_x,
\quad
b=0
\end{gather}
evaluated on the solution space of the $b$-family system. 
Each of these characteristic functions $(P^m,P^u)$ 
has the form of a point symmetry generator \eqref{pointsymm}. 

The characteristic equation for conservation laws 
of the $b$-family system \eqref{bfam-sys-m}--\eqref{bfam-sys-u} is given by 
\begin{equation}\label{bfam-sys-conslaw}
D_t T + D_x X = Q^mG^m + Q^uG^u,
\quad
G^m= m_x + u^{-1}(m_t +bu_x m), 
\quad
G^u= u_{xx}-u+m
\end{equation}
with the multiplier 
\begin{equation}\label{bfam-sys-Q}
Q=(Q^m,Q^u),
\quad
Q^m = E_{m}(X),
\quad
Q^u = E_{u_{x}}(X)
\end{equation}
where $X$ is a function of $t,x,u,u_x,m$, and $t$-derivatives of $u,u_x,m$. 
Consequently, 
multipliers $Q$ are also functions only of $t,x,u,u_x,m$, 
and $t$-derivatives of $u,u_x,m$. 
Since the $b$-family system is of second-order, 
its low-order conservation laws are defined by 
multipliers with a first-order form
\begin{equation}\label{bfam-sys-lowordQ}
Q^m(t,x,u,u_x,m),
\quad
Q^u(t,x,u,u_x,m)
\end{equation}
where $u_t$ is excluded because the only second-order leading derivative 
in the system \eqref{bfam-sys-m}--\eqref{bfam-sys-u} is $u_{xx}$ 
which cannot be obtained from $u_t$ by differentiations. 
The corresponding low-order conserved currents will then have the form 
\begin{equation}\label{bfam-sys-lowordTX}
\Phi=(T(t,x,u,u_x,m),X(t,x,u,u_x,m)) . 
\end{equation}
Thus, the low-order conservation laws admitted by the $b$-family system 
is a strict subset of the low-order conservation laws \eqref{bfam-lowordTX}
admitted by the $b$-family equation \eqref{bfam-eqn}. 

The determining system \eqref{adjsymmdeteq} and \eqref{helmholtzeq} 
for low-order conservation laws \eqref{bfam-sys-lowordTX} 
of the $b$-family system \eqref{bfam-sys-m}--\eqref{bfam-sys-u} 
consists of only the adjoint-symmetry equations
\begin{gather}
( Q^u - Q^m - D_t(u^{-1}Q^m) + u^{-1}bu_x Q^m )|_\E=0 , 
\label{bfam-sys-m-adjsymmeq}\\
( -\Delta Q^u - bu^{-1}D_x(mQ^m) -u^{-2}(bu_xm+m_t)Q^m )|_\E=0 
\label{bfam-sys-u-adjsymmeq}
\end{gather}
since the Helmholtz equations \eqref{helmholtzeq} 
can be shown to hold identically 
due to $Q$ not containing any $t$-derivatives of $u$. 
Hence, 
low-order adjoint-symmetries of the $b$-family system 
are the same as low-order multipliers. 

If $Q^u$ is eliminated in terms of $Q^m$ through equation \eqref{bfam-sys-m-adjsymmeq}, 
then equation \eqref{bfam-sys-u-adjsymmeq} simplifies to 
the adjoint-symmetry determining equation \eqref{bfam-adjsymmeq} 
for the $b$-family equation, 
with $Q=u^{-1}Q^u|_{m=u-u_{xx}}$. 
In particular, 
the low-order multipliers \eqref{bfam-Q-b<>0}--\eqref{bfam-Q-b=3} 
admitted by the $b$-family equation correspond to the multipliers
\begin{gather}
Q^m_1 = um^{-1+1/b},
\quad
Q^u_1 = 0,
\quad
b\neq 0
\label{bfam-sys-Qs-b<>0}\\
Q^m_2 = u, 
\quad
Q^u_2 = (1-b)u_x,
\quad
b\neq 1
\label{bfam-sys-Qs-b<>1}\\
Q^m_3 = u^2,
\quad
Q^u_3 = u_t,
\quad 
b=2
\label{bfam-sys-1stordQs-b=2}\\
Q^m_4= u^2m+\tfrac{1}{2}u(u^2-u_x^2)-uu_{tx},
\quad
Q^u_4= \tfrac{1}{2}u_x(u_x^2-u^2)+(u_t-uu_x)m+um_t-u_{ttx}, 
\quad
b=2
\label{bfam-sys-higherordQs-b=2}\\
Q^m_5 = u^2m+\tfrac{1}{2}u^3 -uu_x^2-uu_{tx},
\quad
Q^u_5 = 2u_x(u_x^2-u^2)+(u_t-2uu_x)m+um_t-u_{ttx}, 
\quad 
b=3
\label{bfam-sys-higherordQs-b=3}\\
Q^m_{6^\pm}= e^{\pm 2x}u, 
\quad
Q^u_{6^\pm}= -2e^{\pm 2x}(u+u_x), 
\quad 
b=3
\label{bfam-sys-lowordQs-b=3}\\
Q^m_7 = u(1+\ln(m)), 
\quad
Q^u_7 = -u_x,
\quad
b=1
\label{bfam-sys-lowordQs-b=1}
\end{gather}
for the $b$-family system. 
Note that only the multipliers \eqref{bfam-sys-Qs-b<>0}, \eqref{bfam-sys-Qs-b<>1}, \eqref{bfam-sys-lowordQs-b=3}, \eqref{bfam-sys-lowordQs-b=1} 
have the low-order form \eqref{bfam-sys-lowordQ}. 
Hence, 
the other multipliers \eqref{bfam-sys-1stordQs-b=2}, \eqref{bfam-sys-higherordQs-b=2}, \eqref{bfam-sys-higherordQs-b=3} 
are higher-order multipliers for the $b$-family system.

This shows that conservation laws of low-order for a single PDE 
can be equivalent to higher-order conservation laws for an equivalent PDE system.  
Nevertheless, the sets of all non-trivial conservation laws (up to equivalence) 
for a single PDE and for an equivalent PDE system will have a one-to-one correspondence.

\subsection{Navier-Stokes equations}
\label{NS-ex}

Our final example is the Navier-Stokes equations \cite{Bat} 
for compressible, viscous fluids in two dimensions
\begin{gather}
\rho_t + \nabla\cdot(\rho\vec u) =0 , 
\label{NS-mass-eqn}\\
(\rho\vec u)_t +\nabla\cdot(\rho \vec u\odot\vec u +p\ \mathbf{I}) = \nabla\cdot\boldsymbol\sigma , 
\label{NS-momentum-eqn}\\
\boldsymbol\sigma = \mu (\nabla\odot\vec u - \tfrac{1}{2}\nabla\cdot\vec u\ \mathbf{I})
\label{NS-stress-eqn}
\end{gather}
where $\rho$ is the density and $\vec u = (u^1,u^2)$ is the velocity, 
which are functions of $t$ and $\vec x=(x,y)$, 
and where $p$ is the pressure, $\mu$ is the viscosity,
and $\boldsymbol\sigma$ is the trace-free stress tensor. 
Here $\odot$ denotes the symmetric product of vectors, 
and $\mathbf{I}$ denotes the $2$x$2$ identity matrix. 
In general, $p$ and $\mu$ are functions of $\rho$. 
Unlike the non-viscous equations, this system has no Hamiltonian structure. 

We will consider the slightly-compressible case, 
in which $\mu$ is constant and $p$ is linear in $\rho$. 
This yields the PDE system 
\begin{subequations}\label{NS-sys}
\begin{gather}
\rho_t + (\rho u^1)_x + (\rho u^2)_y =0 , 
\label{NS-rho-eqn}\\
u^1_t + u^1u^1_x + u^2u^1_y = (1/\rho)(-\kappa \rho_x + \mu(u^1_{xx} + u^1_{yy})), 
\label{NS-u1-eqn}\\
u^2_t + u^1u^2_x + u^2u^2_y = (1/\rho)(-\kappa \rho_y + \mu(u^2_{xx} + u^2_{yy})), 
\label{NS-u2-eqn}\\
\mu=\const\neq 0, 
\quad
\kappa=\const\neq 0 .
\label{NS-slightlycompr}
\end{gather}
\end{subequations}
There are many different sets of leading derivatives for this system \eqref{NS-sys}. 
For instance, 
$\{\rho_t,u^1_t,u^2_t\}$, 
$\{\rho_x,u^1_{xx},u^2_{xx}\}$, 
$\{\rho_y,u^1_{yy},u^2_{yy}\}$
are each a set of leading derivatives. 
Consequently, the system \eqref{NS-sys} is normal. 

The determining system \eqref{symmdeteq} 
for infinitesimal symmetries $\X=P^\rho\p_\rho+P^{u^1}\p_{u^1}+P^{u^2}\p_{u^2}$ 
of the Navier-Stokes system \eqref{NS-sys} is given by 
\begin{subequations}\label{NS-symmeq}
\begin{gather}
\big( D_tP^\rho + D_x(u^1P^\rho + \rho P^{u^1}) +D_y(u^2P^\rho +\rho P^{u^2}) \big)|_\E=0 , 
\label{NS-rho-symmeq}\\
\begin{aligned}
& \big( D_tP^{u^1} + u^1D_xP^{u^1} + u^2D_yP^{u^1} + u^1_x P^{u^1} + u^1_y P^{u^2} + \kappa D_x(\rho^{-1}P^\rho) 
\\&\qquad
-\mu \rho^{-1}(D_x{}^2P^{u^1} + D_y{}^2P^{u^1}) +\mu \rho^{-2}(u^1_{xx} + u^1_{yy})P^\rho \big)|_\E=0 , 
\end{aligned}
\label{NS-u1-symmeq}\\
\begin{aligned}
& \big( D_tP^{u^2} + u^1D_xP^{u^2} + u^2D_yP^{u^2} + u^2_x P^{u^1} + u^2_y P^{u^2} + \kappa D_y(\rho^{-1}P^\rho) 
\\&\qquad
-\mu \rho^{-1}(D_x{}^2P^{u^2} + D_y{}^2P^{u^2}) +\mu \rho^{-2}(u^2_{xx} + u^2_{yy})P^\rho \big)|_\E=0 .
\end{aligned}
\label{NS-u2-symmeq}
\end{gather}
\end{subequations}
Off of the solution space $\E$, 
these equations \eqref{NS-symmeq} take the form 
\begin{equation}\label{NS-symmeq-offE}
\begin{aligned}
& \delta_{(P^\rho,P^{u^1},P^{u^2})} G^\rho = R_P^\rho(G^\rho,G^{u^1}, G^{u^2}),
\\
& \delta_{(P^\rho,P^{u^1},P^{u^2})} G^{u^1} = R_P^{u^1}(G^\rho,G^{u^1}, G^{u^2}),
\\
& \delta_{(P^\rho,P^{u^1},P^{u^2})} G^{u^2}= R_P^{u^2}(G^\rho,G^{u^1}, G^{u^2}),
\end{aligned}
\end{equation}
as given by equation \eqref{deteqoffE}, 
with
\begin{equation}
\begin{aligned}
& G^\rho{} = \rho_t + (\rho u^1)_x + (\rho u^2)_y , 
\\
& G^{u^1} = u^1_t + u^1u^1_x + u^2u^1_y +(1/\rho)(\kappa \rho_x - \mu(u^1_{xx} + u^1_{yy})) , 
\\
& G^{u^2} = u^2_t + u^1u^2_x + u^2u^2_y + (1/\rho)(\kappa \rho_y - \mu(u^2_{xx} + u^2_{yy})) . 
\end{aligned}
\label{NS-G}
\end{equation}

A direct calculation of characteristic functions $P=(P^\rho,P^{u^1},P^{u^2})$ 
for point symmetries yields 
\begin{gather}
P_1 = (-\rho_x,-u^1_x,-u^2_x),
\quad
P_2 = (-\rho_y,-u^1_y,-u^2_y),
\quad
P_3 = (-\rho_t,-u^1_t,-u^2_t),
\\
P_4 = (-t\rho_x,1-tu^1_x,-tu^2_x),
\quad
P_5 = (-t\rho_y,-tu^1_y,1-tu^2_y),
\\
P_6 = (y\rho_x-x\rho_y,yu^1_x-xu^1_y-u^2,yu^2_x-xu^2_y+u^1), 
\\
P_7 = (-t\rho_t-x\rho_x-y\rho_y-\rho,-tu^1_t-xu^1_x-yu^1_y,-tu^2_t-xu^2_x-yu^2_y) ,
\end{gather}
with
\begin{gather}
R_{P_1} = 
\begin{pmatrix}
-D_x & 0 & 0\\
0 & -D_x & 0\\
0 & 0 & -D_x 
\end{pmatrix},
R_{P_2} = 
\begin{pmatrix}
-D_y & 0 & 0\\
0 & -D_y & 0\\
0 & 0 & -D_y 
\end{pmatrix},
R_{P_3} = 
\begin{pmatrix}
-D_t & 0 & 0\\
0 & -D_t & 0\\
0 & 0 & -D_t 
\end{pmatrix}, 
\label{NS-RP-trans}
\\
R_{P_4} = 
\begin{pmatrix}
-tD_x & 0 & 0\\
0 & -tD_x & 0\\
0 & 0 & -tD_x 
\end{pmatrix},
R_{P_5} = 
\begin{pmatrix}
-tD_y & 0 & 0\\
0 & -tD_y & 0\\
0 & 0 & -tD_y 
\end{pmatrix},
\label{NS-RP-boost}
\\
R_{P_6} = 
\begin{pmatrix}
yD_x-xD_y & 0 & 0\\
0 & yD_x-xD_y & -1\\
0 & 1 & yD_x-xD_y 
\end{pmatrix},
\label{NS-RP-rot}
\\
R_{P_7} = 
\begin{pmatrix}
-tD_t-xD_x-yD_y -2& 0 & 0\\
0 & -tD_t -xD_x-yD_y -1& \\
0 & 0 & -tD_t -xD_x-yD_y -1
\end{pmatrix}
\label{NS-RP-scal}
\end{gather}
where $R_P= (R_P^\rho,R_P^{u^1},R_P^{u^2})$ is the matrix 
defined by factoring out the column vector $(G^\rho,G^{u^1}, G^{u^2})^\t$ 
in equation \eqref{NS-symmeq-offE}. 

From this calculation, 
it follows that all point symmetries of the Navier-Stokes system \eqref{NS-sys} 
are generated by 
translations in $x,y,t$
\begin{equation}
\X_1 = \p_x, 
\quad
\X_2 = \p_y, 
\quad
\X_3 = \p_t, 
\label{NS-trans-symms}
\end{equation}
Galilean boosts with respect to $x,y$
\begin{equation}
\X_4 = \p_{u^1} +t\p_x,
\quad
\X_5 = \p_{u^2} +t\p_y,
\label{NS-boost-symms}
\end{equation}
a rotation 
\begin{equation}
\X_6 = -u^2\p_{u^1} +u^1\p_{u^2} -y\p_x + x\p_y,
\label{NS-rot-symm}
\end{equation}
and a scaling
\begin{equation}
\X_7 = -\rho\p_{\rho} +t\p_t +x\p_x + y\p_y . 
\label{NS-scal-symm} 
\end{equation}
Their respective group actions consist of 
\begin{gather}
 x\rightarrow x+\epsilon,
\label{NS-pt1}
\\
y\rightarrow y + \epsilon, 
\label{NS-pt2}
\\
t\rightarrow t+\epsilon,
\label{NS-pt3}
\\
x\rightarrow x +\epsilon t, 
\quad
u^1\rightarrow u^1 + \epsilon 
\label{NS-pt4}
\\
y\rightarrow y +\epsilon t, 
\quad
u^2\rightarrow u^2 + \epsilon 
\label{NS-pt5}
\\
\begin{aligned}
& x\rightarrow \cos(\epsilon)x -\sin(\epsilon)y, 
\quad
y\rightarrow \cos(\epsilon)y +\sin(\epsilon)x, 
\\&\qquad
u^1\rightarrow \cos(\epsilon)u^1 -\sin(\epsilon)u^2, 
\quad
u^2\rightarrow \cos(\epsilon)u^2 +\sin(\epsilon)u^1, 
\end{aligned}
\label{NS-pt6}
\\
x\rightarrow e^\epsilon x, 
\quad
y\rightarrow e^\epsilon y, 
\quad
t\rightarrow e^\epsilon t, 
\quad
\rho \rightarrow e^{-\epsilon}\rho
\label{NS-pt7}
\end{gather}
in terms of a parameter $\epsilon$. 

To formulate the characteristic equation for conservation laws of the Navier-Stokes system \eqref{NS-sys}, 
it is simplest to use the solved form in terms of the set of leading derivatives 
$\{\rho_t,u^1_t,u^2_t\}$. 
This yields 
\begin{equation}\label{NS-conslaw}
D_t T + D_x X +D_y Y = (Q^\rho,Q^{u^1},Q^{u^2})(G^\rho,G^{u^1}, G^{u^2})^\t
\end{equation}
with the multiplier $Q=(Q^\rho,Q^{u^1},Q^{u^2})$ given by 
\begin{equation}\label{NS-QT-rel}
Q^\rho = E_{\rho}(T), 
\quad
Q^{u^1} = E_{u^1}(T),
\quad
Q^{u^2} = E_{u^2}(T)
\end{equation}
where $T$ is a function of $t,x,y,\rho,u^1,u^2$, and $x,y$-derivatives of $\rho,u^1,u^2$,
with all $t$-derivatives being eliminated from $T$ 
through the PDEs in the system \eqref{NS-sys}.  
Consequently, 
multipliers $Q=(Q^\rho,Q^{u^1},Q^{u^2})$ are also functions only of 
$t,x,y,\rho,u^1,u^2$, and $x,y$-derivatives of $\rho,u^1,u^2$. 
Low-order conservation laws are defined by multipliers with the first-order form
\begin{equation}\label{NS-lowordQ}
Q=Q(t,x,y,\rho,u^1,u^2,u^1_x,u^1_y,u^2_x,u^2_y). 
\end{equation}
The corresponding conserved currents $\Phi= (T,X,Y)$ will have the form 
\begin{equation}\label{NS-lowordPhi}
\Phi=\Phi(t,x,y,\rho,u^1,u^2,u^1_x,u^1_y,u^2_x,u^2_y) 
\end{equation}
as determined by equations \eqref{NS-conslaw} and \eqref{NS-QT-rel}. 
For multipliers and currents of this form, 
it is straightforward to show that their dependence on derivatives of $u^1,u^2$
must be at most linear, 
by splitting the equations with respect to derivatives of $\rho,u^1,u^2$, 
and using the linearity of the Navier-Stokes system \eqref{NS-sys} 
in second-order derivatives of $u^1,u^2$. 

The determining system \eqref{adjsymmdeteq} and \eqref{helmholtzeq} 
for all low-order conservation laws \eqref{NS-lowordPhi} consists of 
the adjoint-symmetry equations 
\begin{subequations}\label{NS-adjsymmeq}
\begin{gather}
\begin{aligned}
& \big( -D_tQ^\rho -u^1D_x Q^\rho -u^2D_y Q^\rho -\kappa \rho^{-1}(D_x Q^{u^1} + D_y Q^{u^2}) 
\\&\qquad
+ \mu\rho^{-2}( (u^1_{xx} + u^1_{yy})Q^{u^1} + (u^2_{xx} + u^2_{yy})Q^{u^2} ) \big)\big|_\E=0 ,
\end{aligned} 
\label{NS-rho-adjsymmeq}\\
\begin{aligned}
& \big( -D_tQ^{u^1} -\rho D_x Q^\rho + u^2_x Q^{u^2} -u^1D_xQ^{u^1} -D_y(u^2Q^{u^1}) 
\\&\qquad
-\mu( D_x{}^2(\rho^{-1}Q^{u^1}) + D_y{}^2(\rho^{-1}Q^{u^1}) ) \big)\big|_\E=0 ,
\end{aligned}
\label{NS-u1-adjsymmeq}\\
\begin{aligned}
& \big( -D_tQ^{u^2} -\rho D_y Q^\rho + u^1_y Q^{u^1} -u^2D_yQ^{u^2} -D_x(u^1Q^{u^2}) 
\\&\qquad
-\mu( D_x{}^2(\rho^{-1}Q^{u^2}) + D_y{}^2(\rho^{-1}Q^{u^2}) ) \big)\big|_\E=0 ,
\end{aligned}
\label{NS-u2-adjsymmeq}
\end{gather}
\end{subequations}
and the Helmholtz equations
\begin{subequations}\label{NS-helmholtzeq}
\begin{gather}
Q^\rho_{u^1_x} =0,
\quad
Q^\rho_{u^1_y} =0, 
\quad
Q^\rho_{u^2_x} =0, 
\quad
Q^\rho_{u^2_y} =0, 
\\
Q^{u^1}_{u^1_x} =0, 
\quad
Q^{u^1}_{u^1_y} =0, 
\quad
Q^{u^1}_{u^2_x} =0, 
\quad
Q^{u^1}_{u^2_y} =0, 
\\
Q^{u^2}_{u^1_x} =0, 
\quad
Q^{u^1}_{u^1_y} =0, 
\quad
Q^{u^1}_{u^2_x} =0, 
\quad
Q^{u^1}_{u^2_y} =0, 
\end{gather}
\end{subequations}
after simplifications. 
Thus, all low-order multipliers consist of adjoint-symmetries that have no
dependence on derivatives of $\rho,u^1,u^2$. 
Because the Navier-Stokes system \eqref{NS-sys} has no Hamiltonian structure, 
there is no correspondence between multipliers and symmetry characteristics. 

A straightforward computation of adjoint-symmetries 
\begin{equation} 
Q=(Q^\rho(t,x,y,u^1,u^2),Q^{u^1}(t,x,y,u^1,u^2),Q^{u^2}(t,x,y,u^1,u^2))
\end{equation}
yields
\begin{gather}
Q_1 = (1,0,0), 
\label{NS-Q1}
\\
Q_2 = (u^1,\rho,0),
\quad
Q_3 = (u^2,0,\rho),
\label{NS-Q2&3}
\\
Q_4= (tu^1-x,t\rho,0),
\quad 
Q_5= (tu^2-y,0,t\rho),
\label{NS-Q4&5}
\\
Q_6= (xu^2-yu^1,-y\rho,x\rho) ,
\label{NS-Q6}
\end{gather}
with 
\begin{gather}
R_{Q_1} = 0, 
\label{NS-RQ1}
\\
R_{Q_2} = 
\begin{pmatrix}
0& -1 & 0\\
-1 & 0 &0 \\
0 & 0 & 0
\end{pmatrix},
\quad
R_{Q_3} = 
\begin{pmatrix}
0& 0 & -1 \\
0 & 0 & 0 \\
-1 & 0 &0 
\end{pmatrix},
\label{NS-RQ2&3}
\\
R_{Q_4}= 
\begin{pmatrix}
0& -t & 0 \\
-t & 0 & 0 \\
0 & 0 &0 
\end{pmatrix},
\quad 
R_{Q_5} = 
\begin{pmatrix}
0& 0 & -t \\
0 & 0 &0 \\
-t & 0 & 0 
\end{pmatrix},
\label{NS-RQ4&5}
\\
R_{Q_6} = 
\begin{pmatrix}
0& y & -x \\
y & 0 &0 \\
-x & 0 & 0 
\end{pmatrix} 
\label{NS-RQ6}
\end{gather}
where $R_Q= (R_Q^\rho,R_Q^{u^1},R_Q^{u^2})$ is the matrix 
defined by factoring out the column vector $(G^\rho,G^{u^1}, G^{u^2})^\t$ 
in the form of the adjoint-symmetry equations off of the solution space $\E$, 
\begin{equation}\label{NS-adjsymmeq-offE}
\begin{aligned}
& \delta^*_{(Q^\rho,Q^{u^1},Q^{u^2})} G^\rho = R_Q^\rho(G^\rho,G^{u^1}, G^{u^2}) ,
\\
& \delta^*_{(Q^\rho,Q^{u^1},Q^{u^2})} G^{u^1} = R_Q^{u^1}(G^\rho,G^{u^1}, G^{u^2}) ,
\\
& \delta^*_{(Q^\rho,Q^{u^1},Q^{u^2})} G^{u^2} = R_Q^{u^2}(G^\rho,G^{u^1}, G^{u^2}) ,
\end{aligned}
\end{equation}
as given by equation \eqref{deteqoffE}. 

For each multiplier \eqref{NS-Q1}--\eqref{NS-Q6}, 
a conserved current \eqref{NS-lowordPhi} can be obtained directly by 
either integration of the characteristic equation \eqref{NS-conslaw}
or use of the scaling formula 
$\Psi_G((-t\rho_t-x\rho_x-y\rho_y-\rho,-tu^1_t-xu^1_x-yu^2_y,-tu^2_t-xu^2_x-yu^2_y),(Q^\rho,Q^{u^1},Q^{u^2}))$
from \propref{PQformula}. 
In particular, the components of the scaling formula are given by 
\begin{equation}
\begin{aligned}
\Psi_G^t & = 
-(t\rho_t+x\rho_x+y\rho_y+\rho)Q^\rho  -(tu^1_t+xu^1_x+yu^1_y)Q^{u^1} -(tu^2_t+xu^2_x+yu^2_y)Q^{u^2}) , 
\\
\Psi_G^x & = 
-((tu^1_t+xu^1_x+yu^1_y)\rho+(t\rho_t+x\rho_x+y\rho_y+\rho)u^1)Q^\rho
-(tu^2_t+xu^2_x+yu^2_y)u^1Q^{u^2}
\\&\qquad
-(\kappa (t\rho_t+x\rho_x+y\rho_y+\rho)\rho^{-1} +(tu^1_t+xu^1_x+yu^1_y)u^1)Q^{u^1}
\\&\qquad
+\mu(tu^1_{tx}+xu^1_{xx}+yu^1_{xy}+u^1_x)\rho^{-1}Q^{u^1} 
+\mu(tu^2_{tx}+xu^2_{xx}+yu^2_{xy}+u^2_x)\rho^{-1}Q^{u^2} 
\\&\qquad
-\mu(tu^1_t+xu^1_x+yu^1_y)D_x(\rho^{-1}Q^{u^1})
-\mu(tu^2_t+xu^2_x+yu^2_y)D_x(\rho^{-1}Q^{u^2}) , 
\\
\Psi_G^y & = 
-((tu^2_t+xu^2_x+yu^2_y)\rho+(t\rho_t+x\rho_x+y\rho_y+\rho)u^2)Q^\rho
-(tu^1_t+xu^1_x+yu^1_y)u^2Q^{u^1}
\\&\qquad
-(\kappa (t\rho_t+x\rho_x+y\rho_y+\rho)\rho^{-1} +(tu^2_t+xu^2_x+yu^2_y)u^2)Q^{u^2}
\\&\qquad
+\mu(tu^1_{ty}+xu^1_{xy}+yu^1_{yy}+u^1_y)\rho^{-1}Q^{u^1} 
+\mu(tu^2_{ty}+xu^2_{xy}+yu^2_{yy}+u^2_y)\rho^{-1}Q^{u^2} 
\\&\qquad
-\mu(tu^1_t+xu^1_x+yu^1_y)D_y(\rho^{-1}Q^{u^1})
-\mu(tu^2_t+xu^2_x+yu^2_y)D_y(\rho^{-1}Q^{u^2}) . 
\end{aligned}
\end{equation}
We obtain, modulo the addition of a trivial current and an overall scaling, 
\begin{align}
&\begin{aligned}
T_1 = \rho, 
\quad
X_1 = \rho u^1, 
\quad
Y_1 = \rho u^2 ;
\end{aligned}
\label{NS-TX-1}\\
&\begin{aligned}
T_2 =  \rho u^1
\quad
X_2 = \rho(\kappa+ (u^1)^2) -\mu u^1_x,
\quad
Y_2 = \rho u^1u^2 -\mu u^1_y ;
\end{aligned}
\label{NS-TX-2}\\
&\begin{aligned}
T_3 =  \rho u^2
\quad
X_3 = \rho u^1u^2 -\mu u^2_x,
\quad
Y_3 = \rho(\kappa+ (u^2)^2) -\mu u^2_y ;
\end{aligned}
\label{NS-TX-3}\\
&\begin{aligned}
T_4 =  (tu^1 -x)\rho,
\quad
X_4 = \rho t(\kappa+ (u^1)^2) -x\rho u^1 -\mu tu^1_x,
\quad
Y_4 = \rho t u^1u^2 -x\rho u^2 -\mu t u^1_y ;
\end{aligned}
\label{NS-TX-4}\\
&\begin{aligned}
T_5 =  (tu^2 -y)\rho,
\quad
X_5 = \rho t u^1u^2 -y\rho u^1 -\mu t u^2_x,
\quad
Y_5 = \rho t(\kappa+ (u^2)^2) -y\rho u^2 -\mu tu^2_y ;
\end{aligned}
\label{NS-TX-5}\\
&\begin{aligned}
& T_6 =  (xu^2 -yu^1)\rho,
\quad
X_6 = (xu^2 -yu^1)\rho u^1 -\kappa y\rho +\mu(yu^1_x-xu^2_x+u^2),
\\&\qquad
Y_6 = (xu^2 -yu^1)\rho u^2 +\kappa x\rho + \mu(yu^1_y-xu^2_y-u^1) .
\end{aligned}
\label{NS-TX-6}
\end{align}
The first conserved current represents conservation of mass. 
The second and third conserved currents represent conservation of momentum,
while the fourth and fifth conserved currents represent conservation of Galilean momentum. 
The sixth conserved current represents conservation of angular momentum. 

We will now study the symmetry properties of these conservation laws 
\eqref{NS-TX-1}--\eqref{NS-TX-6}. 
Consider the vector space of conserved currents
\begin{equation}\label{NS-gencurrent}
T = a_1 T_1 +a_2 T_2 +a_3 T_3 +a_4 T_4 +a_5 T_5 +a_6 T_6,
\quad
a_i=\const,
\end{equation}
and the algebra of point symmetries
\begin{equation}\label{NS-gensymm}
\X = c_1\X_1+c_2 \X_2+c_3 \X_3 +c_4 \X_4 +c_5 \X_5 +c_6 \X_6 +c_7 \X_7, 
\quad
c_j=\const . 
\end{equation}
A conservation law $(D_tT+D_xX+D_y Y)|_\E=0$ is homogeneous under the symmetry $\X$ 
iff condition \eqref{invcond} is satisfied, 
where the characteristic of the symmetry generator is given by 
$P= c_1P_1+c_2P_2+c_3 P_3 +c_4 P_4+c_5 P_5 +c_6 P_6 +c_7 P_7$, 
and the multiplier for the conservation law is given by 
$Q= a_1Q_1+a_2Q_2 +a_3Q_3 +a_4Q_4 +a_5Q_5 +a_6 Q_6$. 
By using equations \eqref{NS-RP-trans}--\eqref{NS-RP-scal} and \eqref{NS-RQ1}--\eqref{NS-RQ6}, 
we find that the condition \eqref{invcond} splits 
into a system of bilinear equations on $c_j$ and $a_i$:
\begin{equation}\label{NS-aceqs}
\begin{aligned}
& 
a_6(2c_7 - \lambda)=0,
\quad
a_5(2c_7 - \lambda) -a_4c_6+a_6c_4=0,
\quad
a_4(2c_7 - \lambda) +a_5c_6-a_6c_5=0,
\\
&
a_3(c_7 - \lambda) -a_2c_6+a_5c_3+a_6c_1=0,
\quad
a_2(c_7 - \lambda) +a_3c_6+a_4c_3-a_6c_2=0,
\\
&
a_1(c_7 - \lambda) +a_2c_4+a_3c_5-a_4c_1-a_5c_2=0.
\end{aligned}
\end{equation}
The solutions for $c_j$ in terms of $a_i$ 
determine the symmetry-homogeneity properties of 
the conserved current $\Phi=(T,X,Y)$ modulo trivial currents. 
By considering the subspaces generated by $\{a_i\}$,
solving the system \eqref{NS-aceqs} in each case, 
and merging the solutions, we get the conditions
\begin{subequations}
\begin{align}
& a_1\neq 0,
\quad
a_2=a_3=a_4=a_5=a_6=0;
\quad
\lambda = c_7
\\
& a_1{}^2 + a_2{}^2 + a_3{}^2\neq 0,
\quad
a_4=a_5=a_6=0;
\quad
\lambda = c_7,
\quad
c_6=0,
\quad
a_2c_4+a_3c_5=0
\\
&\begin{aligned}
& a_1{}^2 + a_2{}^2 + a_3{}^2+ a_4{}^2+ a_5{}^2+ a_6{}^2\neq 0,
\quad
a_1a_6-a_2a_5+a_3a_4=0;
\\&\quad
\lambda = 2c_7\neq 0,
\quad
a_2c_6+a_3c_7-a_5c_3-a_6c_1=0,
\quad
a_2c_7-a_3c_6-a_4c_3+a_6c_2=0,
\\&\quad
a_4c_6-a_6c_4=0,
\quad
a_5c_6-a_6c_5=0 
\end{aligned}
\\
&\begin{aligned}
& a_1{}^2 + a_2{}^2 + a_3{}^2+ a_4{}^2+ a_5{}^2+ a_6{}^2\neq 0;
\\&\quad
\lambda = 0,
\quad
c_7=0, 
\quad
a_2c_6-a_5c_3-a_6c_1=0,
\quad
a_3c_6+a_4c_3-a_6c_2=0,
\\&\quad
a_4c_6-a_6c_4=0,
\quad
a_5c_6-a_6c_5=0 
\end{aligned}
\end{align}
\end{subequations}
Hence, we conclude the following.

(1) The symmetry properties of the vector space 
$a_1\Phi_1 +a_2\Phi_2 +a_3\Phi_3 +a_4\Phi_4 +a_5\Phi_5 +a_6\Phi_6+\Phi_\triv$ 
for arbitrary $a_i$ consist of 
invariance under $a_5\X_1-a_4\X_2 -a_6\X_3$ 
and $a_2\X_1+a_3\X_2 +a_4\X_4+a_5\X_5+a_6\X_6$. 

(2) The only additional symmetry properties consist of:
(i) invariance of the subspace $a_1\Phi_1 +\Phi_\triv$ 
under $\X_1$, $\X_2$, $\X_3$, $\X_4$, $\X_5$, $\X_6$;
(ii) invariance of the subspace $a_1\Phi_1 +a_2\Phi_2 +a_3\Phi_3 +\Phi_\triv$
under $\X_1$, $\X_2$, $\X_3$, $a_3\X_4-a_2\X_5$;
(iii) homogeneity of the subspaces $a_1\Phi_1 +\Phi_\triv$ 
and $a_1\Phi_1 +a_2\Phi_2 +a_3\Phi_3 +\Phi_\triv$ 
under $\X_7$ with $\lambda =1$;
(iv) homogeneity of the projective subspace 
$a_1\Phi_1 +a_2\Phi_2 +a_3\Phi_3 +a_4\Phi_4 +a_5\Phi_5 +a_6\Phi_6+\Phi_\triv$, 
$a_1a_6-a_2a_5+a_3a_4=0$, 
under $a_1\X_1 +a_2\X_3+a_4\X_7$ with $\lambda = 2a_4$,
and $a_1\X_2 +a_3\X_3+a_5\X_7$ with $\lambda = 2a_5$,
and $a_3\X_1 -a_2\X_2+a_6\X_7$ with $\lambda = 2a_6$. 

From these properties, it follows that
each of the conservation laws \eqref{NS-TX-1}--\eqref{NS-TX-6} 
is homogeneous under the scaling symmetry $\X_7$. 
Additionally, 
all six conservation laws are are invariant under the time-translation symmetry $\X_3$,
while the mass conservation law \eqref{NS-TX-1} 
and the momentum conservation laws \eqref{NS-TX-2}--\eqref{NS-TX-3} 
are invariant under the space-translation symmetries $\X_1,\X_2$. 
Also, the Galilean momentum conservation laws \eqref{NS-TX-4}--\eqref{NS-TX-5} 
are invariant under both of the Galilean boost symmetries $\X_4,\X_5$,
and the angular momentum conservation law \eqref{NS-TX-6} is invariant under
the rotation symmetry $\X_6$.

\section{Concluding remarks}
\label{remarks}

If a normal PDE system is an Euler-Lagrange system or a Hamiltonian system, 
then there is a direct correspondence between conservation laws 
and variational symmetries or Hamiltonian symmetries. 
In these cases, 
symmetry invariance of conservation laws is connected to 
abelian subalgebras in the symmetry algebra of the PDE system,
as illustrated by the examples in \secref{gKdV-ex} and \secref{bfam-ex}. 
We will explore this connection in detail elsewhere. 

In contrast, 
when a normal PDE system does not have any Lagrangian or Hamiltonian formulation, 
conservation laws instead correspond to adjoint-symmetries that satisfy conditions for being multipliers,
and there is no obvious relationship between 
symmetry invariance of conservation laws 
and properties of the symmetry algebra of the PDE system. 
However, it may still be fruitful to look for mappings 
from symmetries and into multipliers, and vice versa, 
in which case the symmetry properties of conservation laws may be related to 
properties of the symmetry algebra itself. 
This is a promising direction for future work.

\section*{Acknowledgements}

S.C. Anco is supported by an NSERC research grant. 
The referee is thanked for bringing to the attention of the authors 
a number of relevant references in the Russian literature.

\end{document}